\documentclass[aps,prb,amsmath,twocolumn,superscriptaddress]{revtex4}

\usepackage{amsmath}
\usepackage{amssymb}

\usepackage{graphicx}
\usepackage{bm} %Uncomment for using bold math
\usepackage{url}

\usepackage{color}

\usepackage{amsmath}
\DeclareMathOperator{\tr}{tr}
\DeclareMathOperator{\Real}{Re}

\DeclareMathOperator{\Imag}{Im}

\usepackage[percent]{overpic}
\usepackage[export]{adjustbox}
\usepackage{stackengine}
\usepackage{appendix}

\usepackage{natbib}
\usepackage{placeins} % put this in your pre-amble
\usepackage{flafter}  % put this in your pre-amble

\begin{document}

\title{Phase coherent electron transport in asymmetric cross-like Andreev interferometers}
\author{Pavel E. Dolgirev}
\affiliation{Department of Physics, Harvard University, Cambridge Massachusetts 02138, USA
}
\author{Mikhail S. Kalenkov}
\affiliation{I.E. Tamm Department of Theoretical Physics, P.N. Lebedev Physical Institute, 119991 Moscow, Russia}
\affiliation{Moscow Institute of Physics and Technology, Dolgoprudny, 141700 Moscow region, Russia}

\author{Andrei E. Tarkhov}
\affiliation{Skolkovo Institute of Science and Technology, Skolkovo Innovation Center, 3 Nobel St., 143026 Moscow, Russia}

\author{Andrei D. Zaikin}
\affiliation{Institut f{\"u}r Nanotechnologie, Karlsruher Institut f{\"u}r Technologie (KIT), 76021 Karlsruhe, Germany}
\affiliation{National Research University Higher School of Economics, 101000 Moscow, Russia}

\date{\today}

\begin{abstract}

We present a detailed theoretical description of quantum coherent electron transport in voltage-biased cross-like Andreev interferometers. Making use of the charge conjugation symmetry encoded in the quasiclassical formalism, we elucidate a crucial role played by geometric and electron-hole asymmetries in these structures. We argue that a non-vanishing Aharonov-Bohm-like contribution to the current $I_S$ flowing in the superconducting contour may develop only in geometrically asymmetric interferometers making their behavior qualitatively
different from that of symmetric devices. The current $I_N$ in the normal contour -- along with $I_S$ -- is
found to be sensitive to phase-coherent effects thereby also acquiring a $2\pi$-periodic dependence
on the Josephson phase. In asymmetric structures this current develops an odd-in-phase contribution originating from electron-hole asymmetry. We demonstrate that both phase dependent currents $I_S$ and $I_N$ can be controlled and manipulated by tuning the applied voltage, temperature and system topology, thus rendering Andreev interferometers particularly important for future applications in modern electronics.
\end{abstract}

\maketitle

\section{Introduction}

An interplay between quantum coherence and non-equilibrium phenomena is an intriguing topic in condensed matter physics. Hybrid metallic heterostructures composed of superconducting (S) and normal (N) terminals constitute an important playground to realize and investigate rich physics associated with the above phenomena. In these systems -- frequently called Andreev interferometers -- long-range quantum coherence is induced due to the superconducting proximity effect, while non-equilibrium conditions can be created by virtue of biasing different terminals with external voltages and/or temperature gradients~\cite{belzig1999quasiclassical,fornieri2017towards,giazotto2006opportunities}. Distinctive electrical and thermal properties of such systems -- including, e.g., large phase-dependent thermoelectric effects~\cite{eom1998phase,dikin2001low,parsons2003reversal,cadden2007charge,shelly2016resolving},  conductance re-entrance~\cite{stoof1996flux, golubov1997coherent}, Aharonov-Bohm-like behavior of SN-rings~\cite{stoof1996flux,golubov1997coherent,nakano1991quasiparticle,petrashov1995phase,courtois1996long} and non-local (or crossed) Andreev reflection~\cite{byers1995probing,deutscher2000coupling,beckmann2004evidence,russo2005experimental,cadden2006nonlocal,golubev2007non,kalenkov2007crossed,golubev2009crossed} -- render them a promising platform for modern electronics and caloritronics.

Yet another remarkable effect is the so-called $\pi$-junction state that can occur in systems with two normal and two superconducting terminals interconnected by normal metallic wires forming a cross. Applying a phase twist $\phi$ to two superconducting terminals of this cross-like Andreev interferometer one induces dc Josephson current $I_S(\phi)$ between these terminals just like in usual SNS junctions  \cite{ZZh,dubos2001josephson,golubov2004current}. Simultaneously biasing two normal terminals with an external voltage $V$ one can modify the electron distribution function in the system, and thereby control the magnitude of $I_S\equiv I_S(\phi,V)$.  At some values of $V$ the supercurrent flowing between S-terminals becomes {\it negative} signaling the $\pi$-junction state \cite{volkov1995af,wilhelm1998mesoscopic,yip1998energy,baselmans1999reversing}.

Recently three of us demonstrated~\cite{dolgirev2018current,dolgirev2019interplay} that the above scenario -- being appropriate for  symmetric cross-like Andreev interferometers -- becomes by far incomplete as soon as the system topology is made {\it asymmetric}. It turns out that in the latter situation the underlying physics becomes much richer being essentially determined by a competition between voltage-dependent (odd in $\phi$) Josephson and (even in $\phi$) Aharonov-Bohm-like  currents flowing in the superconducting contour. This trade-off may have a drastic impact on the current-phase relation $I_S(\phi)$ in voltage-biased Andreev interferometers resulting in a novel $(I_0,\phi_0)$-junction state~\cite{dolgirev2018current}, predicted to occur at low $T$ and high enough $eV$ exceeding an effective Thouless energy of our device. This state is characterized by coherent $2\pi$-periodic oscillations of $I_S$ as a function of $\phi$ shifted  from the origin by the phase $\phi_0(V)$ that can take {\it any value}, thus being in general different from zero or $\pi$. 

\begin{figure}
    \centering
    \includegraphics[width=0.8\linewidth]{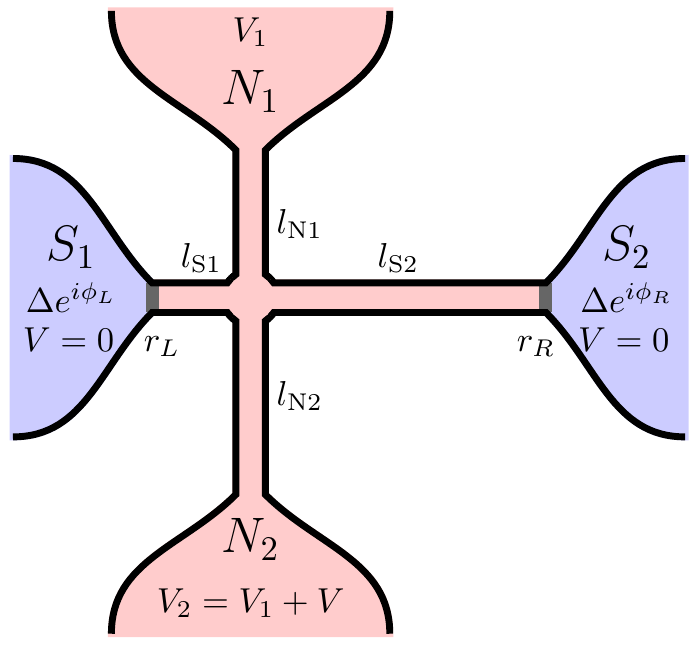}
    \caption{(Color online) Asymmetric cross-like Andreev interferometer under consideration.}
    \label{fig:geom}
\end{figure}

It should be emphasized that asymmetric topology of cross-like Andreev interferometers plays a crucial role for this effect: With the aid of simple charge-conjugation symmetry arguments to be outlined below one can demonstrate that by making the interferometer in Fig.~\ref{fig:geom} symmetric in at least one of the two contours (either normal or superconducting) one totally suppresses the Aharonov-Bohm contribution to $I_S$, hence, getting back to the physical picture \cite{volkov1995af,wilhelm1998mesoscopic,yip1998energy,baselmans1999reversing} with $\phi_0(V)=0,\pi$.

Here we will argue that the physics of asymmetric cross-like interferometers is actually even richer than that discussed in our previous studies~\cite{dolgirev2018current,dolgirev2019interplay}. In particular, it turns out that the current $I_N$ flowing between the two normal terminals of such interferometer -- similarly to $I_S$ -- exhibits proximity induced coherent $2\pi$-periodic oscillations as a function of the superconducting phase difference $\phi$. The function $I_N(\phi)$ is in general neither even nor odd, i.e. it consists of both even and odd in $\phi$ harmonics. While the first of these contributions ($\propto \cos \phi$) to the current $I_N$ can be interpreted in terms of the Aharonov-Bohm-like effect, the second one ($\propto \sin\phi$) is  much more tricky as it obviously cannot have anything to do with the Josephson current. Below we will demonstrate that the physical origin of the latter contribution to $I_N$ is directly related to 
electron-hole asymmetry generated due to the mechanism of sequential  Andreev reflections at different NS-interfaces~\cite{kalenkov2017large}. 
 
The structure of the paper is as follows. In Sec.~\ref{sec:model}, we introduce the system under consideration, describe the quasiclassical formalism used throughout our paper and elucidate the charge-conjugation symmetry properties of this formalism important for our further considerations. In Sec.~\ref{sec:analytics} we focus our analysis on the limit of highly resistive NS interfaces, in which case it is possible to obtain a full analytic solution of our problem.
 Our key results and the corresponding discussion are formulated in Sec.~\ref{sec:results}. In Sec.~\ref{sec:Conclusions}, we briefly summarize our findings. Further technical details are relegated to appendices.

\section{The model and basic formalism}
\label{sec:model}

Below we will consider cross-like Andreev interferometers schematically depicted in Fig.~\ref{fig:geom}.
The system consists of two normal-metal diffusive wires of total lengths $l_{N1}+l_{N2}$ and 
$l_{S1}+l_{S2}=L$ connected between each other in the form of a cross, and attached respectively to two normal and two superconducting terminals. We will  address a general case of asymmetric Andreev interferometers with $l_{N1} \neq l_{N2}$ and $l_{S1} \neq l_{S2}$ which demonstrate a variety of quantum coherent effects some of which do not occur in
symmetric configurations.  Electrostatic potentials of both S-terminals are set equal to zero $V=0$, while the potentials of the normal terminals are denoted as $V_1$ and $V_2$. These N-terminals are biased by an external voltage $V$ implying $V_2=V_1+V$. The superconducting order parameter of the left and right S-terminals is chosen to be respectively $\Delta \exp (i\phi_L)$ and $\Delta \exp (i\phi_R)$. The value of the phase difference between these terminals $\phi =\phi_L-\phi_R$ can easily be controlled by an external magnetic flux inserted inside a superconducting loop. 

Obviously, electric current can flow between S-terminals (superconducting contour) as well as between N-terminals (normal contour) being dependent on external bias $V$, temperature $T$ and phase difference $\phi$. The task at hand is to determine the distribution of voltages and electric currents in our structure in the presence of long-range quantum coherent effects, and to demonstrate the importance of geometric and electron-hole asymmetries in our problem.

\subsection{Quasiclassical formalism}
We will adopt the standard quasiclassical formalism \cite{belzig1999quasiclassical} aimed at describing
non-equilibrium quantum properties of hybrid metallic structures like the one in Fig.~\ref{fig:geom}. The quasiclassical Green's functions in each metallic wire are represented with the aid of a $4\times 4$-matrices in the Keldysh-Nambu space composed of retarded ($\hat{G}^R$), advanced ($\hat{G}^A$) and Keldysh ($\hat{G}^K$) functions
\begin{equation}
\check{G}(\epsilon,\bm r) = \begin{pmatrix}
\hat{G}^R & \hat{G}^K\\\hat{0}
& \hat{G}^A
\end{pmatrix}.
\end{equation}
This matrix Green's function obeys the normalization condition $\check{G}^2 = \check{1}$ and satisfies the Usadel equation
\begin{equation}
D\nabla \left(\check{G} \nabla \check{G} \right) = \left[ -i\epsilon \hat \tau_z,\check{G} \right],
\label{eqn:Usadel}
\end{equation}
where $D$ stands for a diffusion coefficient and $\hat \tau_z$ is the Pauli matrix in the Nambu space.

In what follows it will be convenient for us to employ the so-called Riccati parameterization~\cite{schopohl1995quasiparticle,schopohl1998transformation}.
For the retarded Green's function it reads
\begin{equation}
\hat{G}^R = \frac{1}{1+\gamma\tilde{\gamma}}\begin{pmatrix}
1 - \gamma\tilde{\gamma} & 2 \gamma\\
2\tilde{\gamma} & \gamma\tilde{\gamma} - 1
\end{pmatrix}.
\end{equation}
A similar representation holds for the advanced Green's function since $\hat{G}^A = -\hat \tau_z ( \hat{G}^R )^+ \hat \tau_z$. The spectral part of the Usadel equation then becomes
\begin{eqnarray}
\Delta \gamma -\frac{2\tilde{\gamma}}{1+\gamma\tilde{\gamma}} (\nabla\gamma)^2 + 2i\epsilon\gamma = 0,\label{eqn:gamma}\\
\Delta \tilde{\gamma} -\frac{2\gamma}{1+\gamma\tilde{\gamma}} (\nabla\tilde{\gamma})^2 + 2i\epsilon\tilde{\gamma} = 0.\label{eqn:gamma_t}
\end{eqnarray}

With the aid of the standard representation for the Keldysh Green's function 
\begin{equation}
\hat{G}^K = \hat{G}^R \hat{F} - \hat{F}\hat{G}^A,\ \hat{F} = f_L + \hat \tau_z f_T,
\end{equation}
the kinetic part of the Usadel equation can be cast to the form
\begin{eqnarray}
\nabla j_L &=& 0,\ j_L = D_L \nabla f_L - {\cal Y} \nabla f_T + j_s f_T, \label{eqn::j_L}\\
\nabla j_T &=& 0,\ j_T = D_T \nabla f_T + {\cal Y} \nabla f_L + j_s f_L.\label{eqn::j_T}
\end{eqnarray}
Here $j_T = \tr(\check G \nabla \check G \hat \tau_z)^K$ and $j_L = \tr(\check G \nabla \check G)^K$ represent the spectral densities of respectively electric and thermal currents,  $f_L$ and $f_T$ are respectively symmetric and anisymmetric parts of the electron distribution function, 
\begin{equation}
j_s = \frac{1}{4} \tr \hat \tau_z \left( \hat{G}^R \nabla \hat{G}^R  - \hat{G}^A \nabla \hat{G}^A  \right)
\end{equation}
stands for the supercurrent density, and the kinetic coefficients $D_L$, $D_T$ and ${\cal Y}$ are defined as
\begin{eqnarray}
D_L &=& \frac{1}{2}  - \frac{1}{4}\tr \hat{G}^R \hat{G}^A, \\ 
D_T &=& \frac{1}{2}  - \frac{1}{4}\tr \hat{G}^R \hat \tau_z \hat{G}^A \hat \tau_z,\\ 
{\cal Y} &=& \frac{1}{4} \tr \hat{G}^R \hat \tau_z \hat{G}^A.
\label{eh}
\end{eqnarray}
Note that the kinetic coefficient (\ref{eh}) explicitly accounts for the presence of the electron-hole asymmetry in our system.

Resolving the Usadel equations one can evaluate the electric current density $j$ in our system defined as
\begin{equation}
j = -\frac{\sigma_N}{2e} \int j_T(\epsilon)d\epsilon,
\end{equation}
where $\sigma_N$ is the Drude conductivity of a normal metal.

\subsection{Boundary conditions}
As usually, the Usadel equation~(\ref{eqn:Usadel}) should be supplemented by proper boundary conditions allowing to
match the Green's functions at all inter-metallic interfaces. Below we will assume that the central node -- the contact between the two normal wires -- is characterized by perfect transmission, meaning that the Green's functions are continuous and that the spectral currents associated with them are conserved. The same applies to the boundaries with the N-terminals: The Green's functions inside the normal-metal wire are continuously matched to the corresponding bulk values $\hat{G}^{R/A}_N = \pm \hat \tau_z$ and
\begin{eqnarray}
f^N_{L/T} &=& \frac{1}{2}\left[\tanh\frac{\epsilon + eV}{2T} \pm \tanh\frac{\epsilon - eV}{2T} \right].
\end{eqnarray}
What remains is to define the boundary conditions at two NS interfaces. Here, we will restrict our analysis to the tunneling limit, i.e. we assume that the transmission of both NS interfaces is small compared to unity. This limit is accounted for by the well-known Kupriyanov-Lukichev (KL) boundary conditions~\cite{kuprianov1988influence}
\begin{align}
& L\check{G} \partial_x \check{G} = \pm \frac{1}{2 r} \Big[ \check{G}_{\rm SC}, \check{G} \Big]
\label{eqn: Main_KL},
\end{align}
where $\check{G}$ is the Green's function in the normal wire, $\check{G}_{\rm SC}$ denotes the bulk Green's function of the corresponding S-terminal with
\begin{gather}
\hat{G}^R_{\rm SC} = 
\dfrac{
\begin{pmatrix}
\epsilon & \Delta e^{i\chi} \\
-\Delta e^{-i\chi} & -\epsilon
\end{pmatrix}
}{\sqrt{(\epsilon + i \delta)^2 - \Delta^2}},
\\
\hat{G}^K_{\rm SC} = \tanh\frac{\epsilon}{2 T} \Big( \hat{G}^R_{\rm SC} - \hat{G}^A_{\rm SC}\Big)
\end{gather}
and phase $\chi$ equals to either $\phi_L$ or $\phi_R$ depending on the terminal. The parameter $r$ is defined as
\begin{equation}
r= \dfrac{\mathcal{A} \sigma_N}{L {\mathcal G}},
\label{r}
\end{equation}
where $\mathcal{A}$ is the interface cross section and ${\mathcal G}$ is the normal-state conductance of the interface. 
Note that within the applicability range of KL boundary conditions (\ref{eqn: Main_KL}) and depending on the relation between ${\mathcal G}$ and the conductance of the normal wire of length $L$, the parameter $r$ can in general take any value both smaller and larger than unity.

\subsection{Symmetry considerations}

Let us define charge-conjugated Green's function as
\begin{equation}
\check G_{c} ( \epsilon, \bm{r})
=
- \hat \tau_1 \check G (\epsilon, \bm{r}) \hat \tau_1.
\label{eqn:charge_conj}
\end{equation}
It is straightforward to verify that the function (\ref{eqn:charge_conj}) represents a solution of the Usadel equation~(\ref{eqn:Usadel}) with inverted signs of both electric and magnetic fields as well as of that of the superconducting phase. This symmetry has important consequences for the charge transport properties of the system under consideration.

Resolving the Usadel equation~\eqref{eqn:Usadel} we determine the charge currents in all four metallic wires as functions of the phase difference $\phi$ and the applied voltages $V_1$ and $V_2$. Making use of Eq.  (\ref{eqn:charge_conj}) one can demonstrate that all currents invert their signs under the transformation $V_1 \leftrightarrow - V_1$, $V_2 \leftrightarrow - V_2$, $\phi \leftrightarrow - \phi$, i.e. we have
\begin{eqnarray}
    I_i (-\phi, -V_1,-V_2) = - I_i(\phi, V_1, V_2), 
\label{symcur}
\end{eqnarray}
where the index $i$ labels the wires $N1$, $N2$, $S1$ and $S2$. 

The electrostatic potentials $V_1$ and $V_2$ as functions of both the phase $\phi$ and the bias voltage $V$ are determined from the current conservation conditions
\begin{gather}
I_{N1}(\phi, V_1, V_2) = I_{N2}(\phi, V_1, V_2) = I_N,
\label{Ncons}
\\
I_{S1}(\phi, V_1, V_2) = I_{S2}(\phi, V_1, V_2) = I_S,
\label{Scons}
\end{gather}
combined with the condition $V_2 - V_1 =V$. Likewise, the currents $I_{S1}$, $I_{S2}$, $I_{N1}$, $I_{N2}$ can also be expressed as functions of $\phi$ and $V$. 

In general all these currents are $2\pi$-periodic functions of $\phi$. Extra geometric symmetries of our structure may enforce higher symmetries for the above currents rendering them, e.g., either purely even or purely odd functions of  $\phi$. In particular, it is instructive to distinguish two special cases: (i) symmetric connectors to S-terminals (implying that $l_{S1} = l_{S2}$ and $r_L=r_R$) and (ii) symmetric connectors to N-terminals ($l_{N1} = l_{N2}$). It follows immediately (see also Appendix~\ref{appendix:Symmetry} for more details) that in both cases (i) and (ii)
the current $I_S$ turns out to be an odd function of $\phi$, i.e.
\begin{equation}
I_S(-\phi) = -I_S(\phi),
\label{i}
\end{equation} 
whereas the current $I_N$ is even in $\phi$,
\begin{equation}
I_N(-\phi) = I_N(\phi).
\label{ii}
\end{equation} 
Hence, for partially symmetric cross-like Andreev interferometers (in both cases (i) and (ii)) the Aharonov-Bohm-like contribution to the current $I_S$ vanishes and we are back to the situation of only 0- or $\pi$-junction states considered in Refs.~\cite{volkov1995af,wilhelm1998mesoscopic,yip1998energy}. On top of that, no odd-in-$\phi$ contribution to $I_N$ can occur in such structures. 

In what follows we will, therefore, address the most general case of fully asymmetric interferometers 
with $l_{N1} \neq l_{N2}$ and  $l_{S1} \neq l_{S2}$.

\section{Highly resistive interfaces: analytic solution}\label{sec:analytics}
Let us now employ the above equations and evaluate the Green's functions for the structure depicted in Fig. 1. As usually,
one can split the problem into spectral, Eqs.~(\ref{eqn:gamma}--\ref{eqn:gamma_t}), and kinetic, Eqs.~(\ref{eqn::j_L}--\ref{eqn::j_T}), parts which can be treated separately. Below in this section, we will stick to 
the limit of sufficiently large values of the parameter $r$ at both NS interfaces and construct a full analytic solution of the problem. 

\subsection{Spectral part}
Let us assume that tunnel barriers at both NS interfaces are sufficiently large and, hence, anomalous correlations penetrating into the normal-metal wires from the superconducting terminals are strongly suppressed. In this case, one can linearize the spectral part of the Usadel equation and get
\begin{eqnarray}
\gamma_i'' + \lambda^2 \gamma_i = 0,%\ \tilde{\gamma}_i'' + \lambda^2 \tilde{\gamma}_i = 0,
\label{linUs}
\end{eqnarray}
where $\lambda^2 = 2i\epsilon/(\mathcal{E}_{\rm Th} L^2)$ and $\mathcal{E}_{\rm Th} \equiv D/L^2$ (with $L=l_{\rm S1}+l_{\rm S2}$) is an effective Thouless energy of our setup. The same equation also holds for $\tilde{\gamma}$. Here and below, we also assume $\mathcal{E}_{\rm Th} \ll \Delta$ enabling us to restrict our analysis to subgap energies $|\epsilon|<\Delta$. 

The boundary conditions take the form
\begin{align*}
&\gamma_{\rm S1}(0) = \gamma_{\rm S2}(0) = \gamma_{\rm N1}(0) = \gamma_{\rm N2}(0) = \gamma_0,
\\
&{\cal A}_{\rm S1} \gamma_{\rm S1}'(0) + {\cal A}_{\rm N2} \gamma_{\rm N2}'(0) ={\cal A}_{\rm S2} \gamma_{\rm S2}'(0) + {\cal A}_{\rm N1} \gamma_{\rm N1}'(0),
\\
&\gamma_{\rm N1}(l_{\rm N1}) = \gamma_{\rm N2}(-l_{\rm N2}) = 0,
\\
&L \gamma_{\rm S1}'(-l_{\rm S1}) = -\frac{\mathcal{F}_L}{2r_L},\quad L \gamma_{\rm S2}'(l_{\rm S2}) = \frac{\mathcal{F}_R}{2r_R}.
\end{align*}
Equations in the first two lines follow directly from the continuity of $\gamma$ and from the spectral current conservation at the central node (with coordinate set equal to zero). The equation in the third line 
implies that anomalous correlations vanish at the boundaries with both N-terminals. Finally, the two equations in the last line just represent KL boundary conditions at the left and right NS interfaces characterized by parameters $r_L$ and $r_R$ respectively (defined in Eq. (\ref{r}) with ${\mathcal G}\equiv {\mathcal G}_{L/R}$). We also choose
\begin{align*}
&\mathcal{F}_R = \frac{-i \Delta e^{i\phi_R}}{\sqrt{\Delta^2 - (\epsilon + i\delta)^2}},\ \mathcal{F}_L = \frac{-i \Delta e^{i\phi_L}}{\sqrt{\Delta^2 - (\epsilon + i\delta)^2}}.
%&\mathcal{\tilde{F}}_R = \frac{i \Delta}{\sqrt{\Delta^2 - (\epsilon + i\delta)^2}},\ \mathcal{\tilde{F}}_L = \frac{i \Delta e^{-i\phi}}{\sqrt{\Delta^2 - (\epsilon + i\delta)^2}}.
\end{align*}

Resolving the linearized Usadel equations with the above boundary conditions, we obtain
\begin{gather}
\label{24}
\gamma_{\mathrm{N1}} = \gamma_0 
\dfrac{\sin \lambda (l_{\mathrm{N1}} - x)}{\sin \lambda l_{\mathrm{N1}}},
\quad
\gamma_{\mathrm{N2}} = \gamma_0 
\dfrac{\sin \lambda (l_{\mathrm{N2}} + x)}{\sin \lambda l_{\mathrm{N2}}},
\\
\gamma_{\mathrm{S1}} = \gamma_0 
\dfrac{\cos \lambda (l_{\mathrm{S1}} + x)}{\cos \lambda l_{\mathrm{S1}}}
-
\dfrac{\mathcal{F}_L}{2 L \lambda r_L}
\dfrac{\sin \lambda x}{\cos \lambda l_{\mathrm{S1}}},
\\
\gamma_{\mathrm{S2}} = \gamma_0 
\dfrac{\cos \lambda (l_{\mathrm{S2}} - x)}{\cos \lambda l_{\mathrm{S2}}}
+
\dfrac{\mathcal{F}_R}{2 L \lambda r_R}
\dfrac{\sin \lambda x}{\cos \lambda l_{\mathrm{S2}}},
\end{gather}
where 
\begin{eqnarray}
\gamma_0 &=& \frac{1}{\cal N}\frac{1}{2 L \lambda} \Big( \frac{\mathcal{F}_L {\cal A}_{\rm S1}}{r_L \cos{\lambda l_{\rm S1}}}  + \frac{\mathcal{F}_R {\cal A}_{\rm S2}}{r_R \cos{\lambda l_{\rm S2}}}\Big),\\
{\cal N} &=&  {\cal A}_{\rm N1} \cot{\lambda l_{\rm N1}} + {\cal A}_{\rm N2} \cot{\lambda l_{\rm N2}} - \notag{}\\
&& - {\cal A}_{\rm S1} \tan{\lambda l_{\rm S1}} - {\cal A}_{\rm S2} \tan{\lambda l_{\rm S2}}.
\label{28}
\end{eqnarray}
Then, for the spectral supercurrent density, one readily finds
%\begin{widetext}
\begin{eqnarray}
&&j_{s,\rm S1}{\cal A}_{\rm S1} 
= 
j_{s,\rm S2}{\cal A}_{\rm S2} 
= 
\frac{{\cal A}_{\rm S1} {\cal A}_{\rm S2}}{L^2 r_R r_L} \sin{\phi}\times\notag{}\\
&&\times\Real{  \left\{ \frac{\Delta^2}{\Delta^2 - (\epsilon+ i\delta)^2} \frac{i}{\lambda \cos{\lambda l_{\rm S1}} \cos{\lambda l_{\rm S2}} }  \frac{1}{\cal N} \right\} }. \label{eqn::spec_dens}
\end{eqnarray}
%\end{widetext}

The above analytic solution of the spectral part of the problem enables one to easily derive the applicability condition for the
linearized Usadel equation (\ref{linUs}). Setting functions $\gamma$ to be much smaller than unity within the normal wires and making use of Eqs. (\ref{24})--(\ref{28}), we arrive at the following conditions
\begin{gather}
r_L \gg \dfrac{\mathcal{A}_{\mathrm{S1}}}{
L\left(
\dfrac{\mathcal{A}_{\mathrm{N1}}}{l_\mathrm{N1}}
+
\dfrac{\mathcal{A}_{\mathrm{N2}}}{l_\mathrm{N2}}
\right)
},
\
r_R \gg \dfrac{\mathcal{A}_{\mathrm{S2}}}{
L\left(
\dfrac{\mathcal{A}_{\mathrm{N1}}}{l_\mathrm{N1}}
+
\dfrac{\mathcal{A}_{\mathrm{N2}}}{l_\mathrm{N2}}
\right)
}.
\label{tunlim}
\end{gather}
Note that depending on the system parameters these conditions may substantially deviate from simple inequalities $r_{R,L} \gg 1$ which one could naively expect to be sufficient in order to linearize the Usadel equations.

\subsection{Kinetic part}

Below, we proceed similarly to the above subsection, and resolve the kinetic equations perturbatively by formally expanding them in $1/(r_ir_j)$, where $i,j=L,R$. In the zeroth order, we have
\begin{eqnarray}
    \nabla j^{(0)}_{L/T} = 0,\ j^{(0)}_{L/T} = \nabla f^{(0)}_{L/T}
\end{eqnarray}
with the boundary conditions
\begin{eqnarray*}
&&f^{(0)}_{L/T}(l_{\rm N1},\epsilon) = f_{L/T}^N (V_1),\ f^{(0)}_{L/T}(-l_{\rm N2},\epsilon) = f_{L/T}^N (V_2);\\
&&j^{(0)}_{L,\rm S1/S2} = j^{(0)}_{T,\rm S1/S2} = 0; \\
&& {\cal A}_{\rm S1} j_{L/T, \rm S1} + {\cal A}_{\rm N2} j_{L/T, \rm N2} = {\cal A}_{\rm S2} j_{L/T, \rm S2} + {\cal A}_{\rm N1} j_{L/T, \rm N1}.
\end{eqnarray*}
Equations in the first line  account for boundaries with both N-terminals, the second line represents
KL boundary conditions at both NS interfaces, whereas the last equation just reflects both electric and energy currents conservation and, hence, it remains valid to all orders. In fact, the condition $j_{L, \rm S1/S2} = 0$ is also valid to all orders at energies $|\epsilon| < \Delta$, since subgap excitations do not contribute to the energy current flowing into the S-terminals. 

We observe that -- to the zeroth order -- functions $f^{(0)}_L$ and $f^{(0)}_T$ depend linearly on the coordinate along the wire in the normal contour, whereas in the wire that belongs to the superconducting contour these functions remain constant equal to
\begin{equation}
f^{(0)}_{L/T}(0) =  \left( \frac{1}{R_{\rm N1}} + \frac{1}{R_{\rm N2}} \right)^{-1} \left[\frac{f_{L/T}^N (V_1)}{ R_{\rm N1}}  + \frac{f_{L/T}^N (V_2)}{ R_{\rm N2}} \right].\label{eqn::zeroth_f_LT}
\end{equation}
Here, $R_{\mathrm{N1}}$ and $R_{\mathrm{N2}}$ are normal-state Drude resistances of the normal wires connected to terminals $N_1$ and $N_2$
\begin{equation}
R_{\mathrm{N1}} = \dfrac{l_{\mathrm{N1}}}{\mathcal{A}_{\mathrm{N1}}\sigma_N},
\quad
R_{\mathrm{N2}} = \dfrac{l_{\mathrm{N2}}}{\mathcal{A}_{\mathrm{N2}}\sigma_N}.
\end{equation}

Note that with the aid of the above zeroth order solution combined with KL boundary conditions one can establish the spectral electric current in the superconducting contour to the next order in parameter $\sim 1/(r_Lr_R)$. Indeed, the latter conditions can be written in the form $j^{(1)}_{T} = \pm \alpha_T / (L r_{L/R}) $ with
\begin{eqnarray}
\alpha_T &=& \Imag\left\{ \frac{\Delta ( e^{-i\phi} \gamma + e^{i\phi} \tilde{\gamma} ) }{\sqrt{\Delta^2 - (\epsilon+i\delta)^2}}  \right\}f^{(0)}_L - \notag{}\\
&&- \Imag\left\{\frac{\Delta  ( e^{-i\phi} \gamma - e^{i\phi} \tilde{\gamma} ) }{\sqrt{\Delta^2 - (\epsilon+i\delta)^2}} \right\} f^{(0)}_T .    
\end{eqnarray}
With this in mind, the electric current conservation condition yields
\begin{eqnarray}
    {\cal A}_{\rm S1} \int d\epsilon j_{T, \rm S1}(\epsilon) = {\cal A}_{\rm S2} \int d\epsilon j_{T, \rm S2}(\epsilon),\label{eqn::current_cons}
\end{eqnarray}
which after some algebra can further be cast to the form 
\begin{widetext}
\begin{eqnarray}
&&\frac{{\cal A}_{\rm S2}}{r^2_R} \int d\epsilon f^{(0)}_T  \Imag{  \left\{\frac{\Delta^2}{\Delta^2 - (\epsilon+i\delta)^2} \frac{i}{\lambda } \frac{\tan{\lambda l_{\rm S2} ({\cal A}_{\rm N1} \cot{\lambda l_{\rm N1}} + {\cal A}_{\rm N2} \cot{\lambda l_{\rm N2}} - {\cal A}_{\rm S1} \tan{\lambda l_{\rm S1}}) + {\cal A}_{\rm S2}}}{{\cal N}} \right\} }+\notag{}\\
&&+2\frac{{\cal A}_{\rm S1} {\cal A}_{\rm S2}}{r_R r_L} \cos{\phi} \int d\epsilon f_T^{(0)}  \Imag{  \left\{ \frac{\Delta^2}{\Delta^2 - (\epsilon+i\delta)^2}\frac{i}{\lambda \cos{\lambda l_{\rm S1}}  \cos{\lambda l_{\rm S2}} }  \frac{1}{{\cal N}} \right\} }+\notag{}\\
&&+\frac{{\cal A}_{\rm S1}}{r^2_L} \int d\epsilon f_T^{(0)}  \Imag{  \left\{ \frac{\Delta^2}{\Delta^2 - (\epsilon+i\delta)^2}\frac{i}{\lambda }  \frac{\tan{\lambda l_{\rm S1} ({\cal A}_{\rm N1} \cot{\lambda l_{\rm N1}} + {\cal A}_{\rm N2} \cot{\lambda l_{\rm N2}} - {\cal A}_{\rm S2} \tan{\lambda l_{\rm S2}}) + {\cal A}_{\rm S1}}}{\cal N} \right\} } =0. \label{eqn::main}
\end{eqnarray}
\end{widetext}
This equation together with the condition $V_2-V_1 = V$ defines electrostatic potentials of both normal terminals $V_1$ and $V_2$ demonstrating that these potentials depend not only on $V$ and $T$, but also on phase difference $\phi$ between the superconducting terminals. The latter dependence clearly illustrates the importance of long-range proximity induced quantum coherence effects spreading not only into the superconducting contour but also into the normal contour, thereby influencing the potentials of both normal terminals. It follows from Eq.~(\ref{eqn::main}) that both electrostatic potentials $V_1$ and $V_2$ depend on $\cos\phi$, thus being even functions of $\phi$. 

The above perturbative analysis of the kinetic equations can be justified if the interface resistances are much larger than the resistances of the corresponding attached normal wires
\begin{equation}
r_L L/l_{\mathrm{S1}}, \
r_R L/l_{\mathrm{S2}} \gg 1.
\end{equation}

Having determined $V_1$ and $V_2$, we are ready to find the electric current in the superconducting contour. It reads 
\begin{equation}
I_S = -\frac{\sigma_N}{2e} \int d\epsilon {\cal A}_{\rm S1} j_{T,\rm S1}(\epsilon),
\end{equation} 
where
\begin{widetext}
\begin{eqnarray}
{\cal A}_{\rm S1} j_{T,\rm S1} &=& f^{(0)}_L 
\frac{{\cal A}_{\rm S1} {\cal A}_{\rm S2}}{L^2 r_R r_L} 
\sin{\phi} \Real{  \left\{ \frac{\Delta^2}{\Delta^2 - (\epsilon+i\delta)^2}\frac{i}{\lambda \cos{\lambda l_{\rm S1}}  \cos{\lambda l_{\rm S2}} }  \frac{1}{{\cal N}} \right\} } +\notag{}\\
&+& f^{(0)}_T 
\frac{{\cal A}_{\rm S1} {\cal A}_{\rm S2}}{L^2 r_R r_L} 
\cos{\phi}  \Imag{  \left\{ \frac{\Delta^2}{\Delta^2 - (\epsilon+i\delta)^2}\frac{i}{\lambda \cos{\lambda l_{\rm S1}}  \cos{\lambda l_{\rm S2}} }  \frac{1}{{\cal N}} \right\} } + \notag{}\\
&+& f^{(0)}_T 
\frac{{\cal A}_{\rm S1}}{L^2 r^2_L}  
\Imag{  \left\{ \frac{\Delta^2}{\Delta^2 - (\epsilon+i\delta)^2}\frac{i}{\lambda }  \frac{\tan{\lambda l_{\rm S1} ({\cal A}_{\rm N1} \cot{\lambda l_{\rm N1}} + {\cal A}_{\rm N2} \cot{\lambda l_{\rm N2}} - {\cal A}_{\rm S2} \tan{\lambda l_{\rm S2}}) + {\cal A}_{\rm S1}}}{\cal N} \right\} }.
\label{eqn::spec_dens_I_S}
\end{eqnarray}
\end{widetext}
The first term in the right-hand side of Eq. (\ref{eqn::spec_dens_I_S}) represents the Josephson contribution, the second term (proportional to both $1/{(r_L r_R)}$ and $\cos\phi$) defines the coherent Aharonov-Bohm-like current, while the last term
has to do with the Andreev conductance of SN interfaces. 

\begin{figure}[hbp!]
\centering\includegraphics[width=1\linewidth]{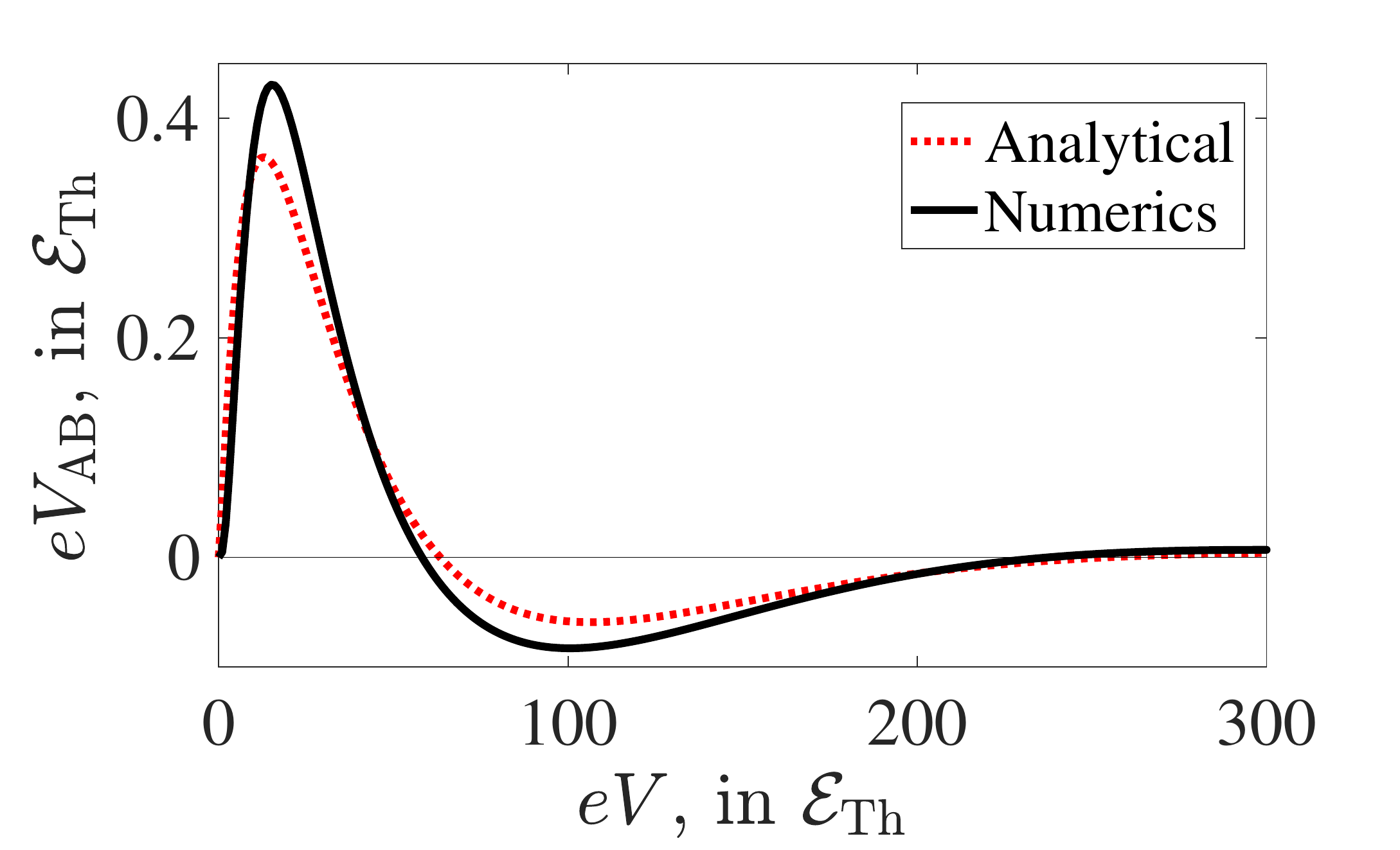}
    \caption{(Color online) The amplitude of even-in-$\phi$ oscillations of electrostatic potentials $V_{1/2}$ as a function of the bias voltage $V$ at $T\mapsto 0$. Numerical curve (solid line) is obtained by solving Eq.~(\ref{eqn::main}). An approximate result (\ref{VAB}) is indicated by the dotted red line. Here we set $r_L = r_R = r\gg1$, $l_{S1}\equiv L-l_{S2}=0.2L$, $l_{N1}=0.3L$, $l_{N2}=0.7L$ and $\Delta = 500{\cal E}_{\rm Th}$. Note that this amplitude does not depend on $r$ in the limit of large $r$.}
    \label{fig:V1}
\end{figure}

As already mentioned above, Eq.~(\ref{eqn::main}) contains only terms depending on $\cos \phi$, while the Josephson contribution proportional to $\sin \phi$ drops out from this equation. In other words, the terms entering the electric current conservation condition, cf. Eq.~\eqref{eqn::current_cons}, represent the combination of $\cos\phi$-dependent (Aharonov-Bohm) and $\phi$-independent (Andreev) contributions. This observation appears to be specific to the chosen cross-like geometry (as suggested, e.g., by Eqs.~(\ref{eqn::spec_dens}) and (\ref{eqn::zeroth_f_LT})) and, furthermore, it only holds in the leading order in $1/(r_ir_j)$. A more detailed numerical analysis indicates that for smaller values of $r_{L,R}$ the $\sin \phi$-harmonic is present and might even play an important role. We also note that Eq.~(\ref{eqn::main}) and Eq.~(\ref{eqn::spec_dens_I_S}) can be combined in a way that allows to expel an explicit dependence on $\cos\phi$ from the expression for $I_S$. In this case, the 
even in $\phi$ contribution to $I_S$ appears implicitly due to the dependencies of potentials $V_1$ and $V_2$ on $\phi$.

% \vskip 1cm

\section{Results and discussion}\label{sec:results}

Let us now explicitly evaluate the distribution of voltages and currents in our cross-like Andreev interferometer.
We first determine electrostatic potentials of the two normal terminals, $V_{1}$ and $V_{2}$, and then evaluate 
the currents in both superconducting and normal contours which depend on these potentials.

\subsection{Electrostatic potentials}

According to Eq.~\eqref{eqn::main} the corresponding Aharonov-Bohm term is proportional to $\sim 1/(r_Lr_R)$, i.e. it has the same order as the other two terms $\sim 1/r_{L}^2$ and $\sim 1/r_{R}^2$. On the other hand,  in the limit $l_{\mathrm{S1}}, l_{\mathrm{S2}} \gg \sqrt{D/\Delta}$ considered here and at high enough voltages $\mathcal{E}_{\mathrm{Th}} \ll e|V_{1,2}| < |\Delta|$ the Aharonov-Bohm contribution becomes exponentially suppressed as $\propto e^{-\sqrt{e|V_{1,2}|/\mathcal{E}_{\mathrm{Th}}}}$ and, hence, it can be treated as a small perturbation.  Then we obtain
\begin{eqnarray}
    V_{1/2} \approx \Bar{V}_{1/2}(V) + V_{\rm AB}(V) \cos\phi ,
\label{V12}
\end{eqnarray}
where 
\begin{eqnarray}
    \Bar{V}_1 \approx -\frac{R_{\rm N1}^2}{R_{\rm N1}^2 + R_{\rm N1}^2} V,\ \Bar{V}_2 \approx \frac{R_{\rm N2  }^2}{R_{\rm N1}^2 + R_{\rm N1}^2} V
\end{eqnarray}
and
\begin{widetext}
\begin{eqnarray}
    eV_{\rm AB} = \frac{4r_Rr_L}{r_R^2 + r_L^2} \frac{(R_{\rm N1} R_{\rm N2})^2}{(R_{\rm N1}^2 + R_{\rm N2}^2)^{3/2}} \sqrt{e|V|{\cal E}_{\rm Th}} \left[ \frac{e^{-\sqrt{e|\Bar{V}_2|/{\cal E}_{\rm Th}}}}{R_{\rm N2}} \sin \sqrt{e|\Bar{V}_2|/{\cal E}_{\rm Th}} - \frac{e^{-\sqrt{e|\Bar{V}_1|/{\cal E}_{\rm Th}}}}{R_{\rm N1}} \sin \sqrt{e|\Bar{V}_1|/{\cal E}_{\rm Th}} \right].
\label{VAB}
\end{eqnarray}
\end{widetext}
The presence of the $\phi$-dependent term in Eq. (\ref{V12}) indicates that electrostatic potentials $V_{1/2}$ are sensitive to proximity-induced long-range quantum coherence in our structure. The magnitude of this coherent contribution to $V_{1,2}$ is controlled by parameter $V_{\rm AB}$ defined in Eq. (\ref{VAB}). Note that this
approximate analytic expression for $V_{\rm AB}$ turns out to be very accurate, as it is demonstrated in Fig.~\ref{fig:V1}. 

\subsection{Current $I_S$ in the superconducting contour}

Following our previous analysis~\cite{dolgirev2018current,dolgirev2019interplay} we can express the current $I_S$ in the superconducting contour in the form
\begin{align}
    &I_S(\phi, V) = I^{\rm even}_S(\phi, V) + I^{\rm odd}_S(\phi, V) \approx\notag{}&\\
    &\approx I_0(V) + I_J(V) \sin\phi + I_{\rm AB}(V)\cos\phi.\label{eqn::I_S_gen}
\end{align}
Here $I_0(V)$ defines a dissipative Andreev-like term entering Eq.~(\ref{eqn::spec_dens_I_S}), while $I_J$  and $I_{\rm AB}$ represent odd in $\phi$ Josephson and even in $\phi$ Aharonov-Bohm-like currents. It follows directly from Eq.~(\ref{eqn::main}) that the last (Aharonov-Bohm) term differs from zero
only for asymmetric structures with both $l_{\rm S1} \neq l_{\rm S2}$ and $l_{\rm N1} \neq l_{\rm N2}$, in accordance with our
general symmetry analysis in Sec IIC.

\begin{figure}[bp!]
\begin{minipage}[h]{1\linewidth}
\center{\includegraphics[width=1\linewidth]{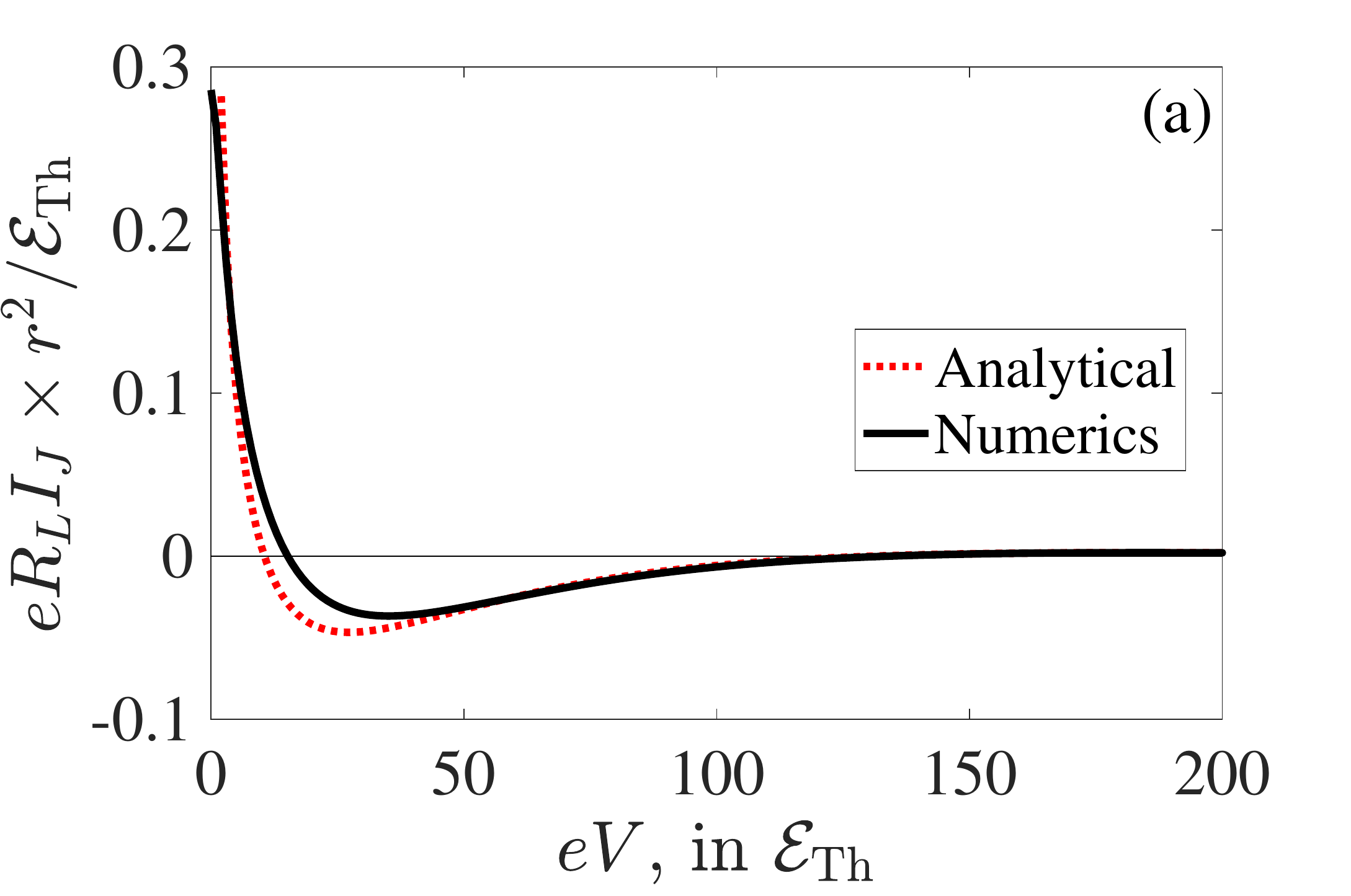}} \\
\end{minipage}
\vfill
\begin{minipage}[h]{1\linewidth}
\center{\includegraphics[width=1\linewidth]{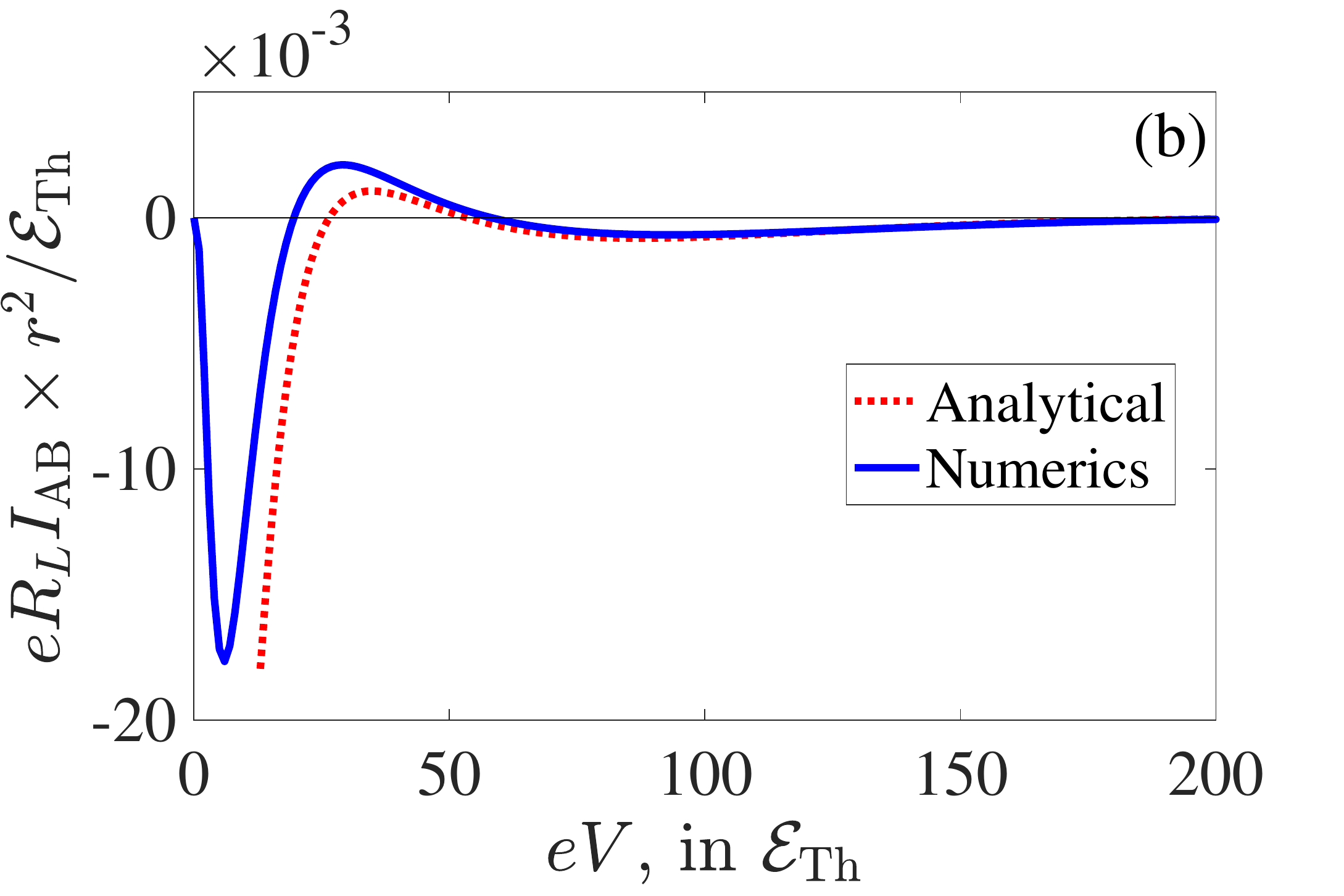}}
\end{minipage}
\caption{(Color online) The amplitudes of odd-in-$\phi$ (a) and even-in-$\phi$ (b) oscillations of the current $I_S$ as functions of $V$ at $T\mapsto 0$. Numerical curves (indicated by solid lines) are obtained by solving Eq.~(\ref{eqn::main}). Analytical results (dotted red lines) correspond to Eq.~(\ref{eqn::I_J_approx}) in (a) and to Eq.~(\ref{eqn::I_AB_approx_eq_r}) in (b).
%%%%%%%%%%%%%%%%%%%%%%%%%%%%%%%%%%%%%%%%%%%%%%%%%% 
\label{fig::I_S}
%%%%%%%%%%%%%%%%%%%%%%%%%%%%%%%%%%%%%%%%%%%%%%%%%%
}
\end{figure}

The results derived in the previous subsection imply that in the particular case of identical NS boundaries with $r_L = r_R = r\gg 1$ we have $I_S(r) \propto 1/r^2$. In this limit, the amplitudes of both odd and even in  $\phi$ oscillations as functions of $V$ are shown in Fig.~\ref{fig::I_S}. We observe that $I_J(V)$ experiences zero-to-$\pi$-junction switching \cite{volkov1995af,wilhelm1998mesoscopic,yip1998energy,baselmans1999reversing} at around $eV \simeq 17{\cal E}_{\rm Th}$ and becomes exponentially suppressed at higher voltages, similarly to the case of fully transparent NS boundaries \cite{dolgirev2018current,dolgirev2019interplay}. Integrating  the supercurrent density in Eq.~(\ref{eqn::spec_dens_I_S}) over energies, at sufficiently high  bias voltages $eV \gg {\cal E}_{\rm Th}$ we get
\begin{eqnarray}
    \frac{eR_LI_J}{{\cal E}_{\rm Th}} &\approx& \frac{1}{r_R r_L} \sum\limits_{i = 1,2} \frac{1/R_{\rm Ni}}{1/R_{\rm N1}+1/R_{\rm N2}}\frac{\Delta^2 }{\Delta^2 - (e\Bar{V}_i)^2} \times\notag{}\\
    && \times e^{-\sqrt{e|\Bar{V}_i|/{\cal E}_{\rm Th}}} \cos{\sqrt{\frac{e|\Bar{V}_i|}{{\cal E}_{\rm Th}}}}.\label{eqn::I_J_approx}
\end{eqnarray}
This formula turns out to be in a good agreement with the numerical solution of Eq.~(\ref{eqn::main}) at $eV \gtrsim 5{\cal E}_{\rm Th}$, cf. Fig.~\ref{fig::I_S}a. 

Just like in the case of transparent SN-boundaries~\cite{dolgirev2018current,dolgirev2019interplay}, here one could expect $I_{\rm AB}(r,V)$ to saturate to some non-zero value $I_{\rm AB}^{\infty}(r)$ at sufficiently high voltages $eV \gg {\cal E}_{\rm Th}$. In contrast to such expectations, in the limit $r \gg 1$ one finds $I_{\rm AB}^{\infty} = 0$, cf. also Fig.~\ref{fig::I_S}b. 
The latter result applies in the leading order in $1/r^2$ and has the same origin as a similar behavior of the coherent contribution to $V_{1/2}$ at large voltages, cf. Fig.~\ref{fig:V1}. Hence, one can expect that $I_{\rm AB}^{\infty}(r) \propto 1/r^4$ for $r\gg 1$.

For amplitude $I_{\rm AB}(V)$, within the voltage interval $\Delta \gg eV \gg {\cal E}_{\rm Th}$, one can derive an expression similar to the one in Eq.~(\ref{eqn::I_J_approx}). Under the condition $l_{\rm S1},\,l_{\rm S2}  \gg \sqrt{D/(eV)}$ we obtain
\begin{eqnarray}
   \frac{eR_LI_{\rm AB}}{{\cal E}_{\rm Th}} &\approx& \frac{r_L^2-r_R^2}{(r_L^2+r_R^2)r_R r_L} \frac{R_{\rm N1}R_{\rm N2}}{R_{\rm N1}+R_{\rm N2}} \times\notag{}\\
    &&  \sum\limits_{i = 1,2}  \frac{(-1)^i }{R_{\rm Ni}} e^{-\sqrt{e|\Bar{V}_i|/{\cal E}_{\rm Th}}} \sin{\sqrt{\frac{e|\Bar{V}_i|}{{\cal E}_{\rm Th}}}}. \label{eqn::I_AB_approx}
\end{eqnarray}
Note that in the particular case $r_R = r_L = r$, the expression~(\ref{eqn::I_AB_approx}) vanishes identically implying that a more accurate treatment is required in this case. The corresponding analysis can be worked out and yields 
\begin{widetext}
\begin{eqnarray}
       \frac{eR_LI_{\rm AB}}{{\cal E}_{\rm Th}} &\approx& -\frac{1}{2r^2} \frac{R_{\rm N1}R_{\rm N2}}{(R_{\rm N1}+R_{\rm N2}) (R_{\rm N1}^2 + R_{\rm N2}^2)} \left[\sum\limits_{i = 1,2}  \frac{(-1)^i }{R_{\rm Ni}} e^{-\sqrt{e|\Bar{V}_i|/{\cal E}_{\rm Th}}} \sin{\sqrt{\frac{e|\Bar{V}_i|}{{\cal E}_{\rm Th}}}}\right] \times \notag{}\\
       &&\times \left\{ \sum\limits_{i,j=1,2} (-1)^j R_{\rm Ni}^2 \,\exp\left(-\sqrt{\frac{e|\Bar{V}_i|}{{\cal E}_{\rm Th}}}\frac{l_{\rm Sj}}{L}\right) \left[\cos\left(\sqrt{\frac{e|\Bar{V}_i|}{{\cal E}_{\rm Th}}}\frac{l_{\rm Sj}}{L}\right) - \sin\left(\sqrt{\frac{e|\Bar{V}_i|}{{\cal E}_{\rm Th}}}\frac{l_{\rm Sj}}{L}\right) \right] \right\}. \label{eqn::I_AB_approx_eq_r}
\end{eqnarray}
\end{widetext}

As far as the temperature dependence of $I_S$ is concerned, we point out that, while the Aharonov-Bohm-like contribution $I_{\rm AB}$ decays as a power law with increasing $T > eV$, the Josephson term $I_{J}$ decays exponentially, thus becoming negligible as compared to $I_{\rm AB}$ in the high-temperature limit $T \gg {\cal E}_{\rm Th}$. In the case of fully transparent NS-interfaces the temperature dependencies of both even and odd in $\phi$ components of $I_S$ have already been studied elsewhere~\cite{dolgirev2018current,dolgirev2019interplay}, therefore we can avoid further details here.

\subsection{Current $I_N$ in the normal contour}

\begin{figure}[htp!]
\begin{minipage}[h]{1\linewidth}
\centering
\def\big{\includegraphics[width=1\linewidth, center]{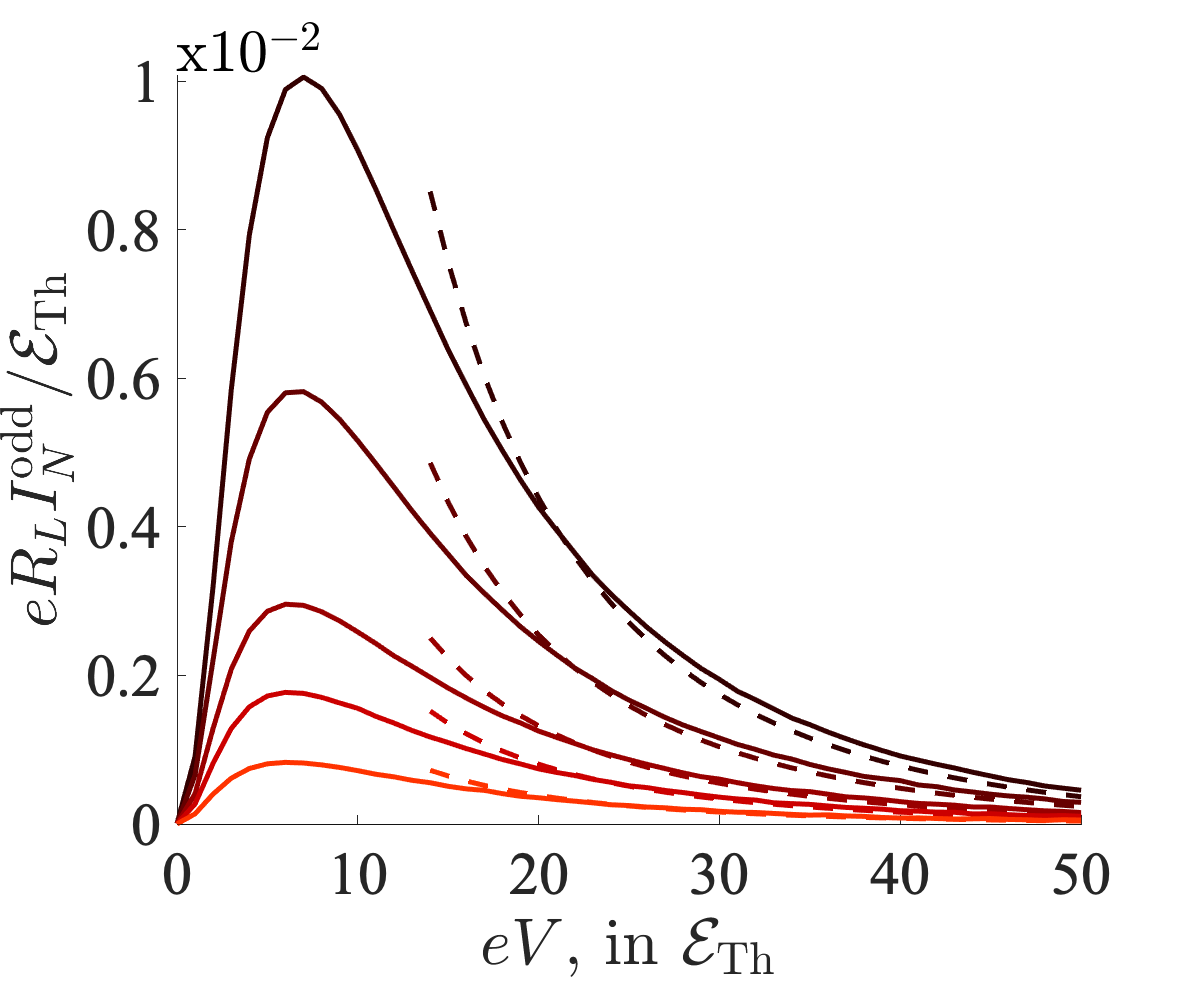}}
\def\little{\includegraphics[height=3.6cm]{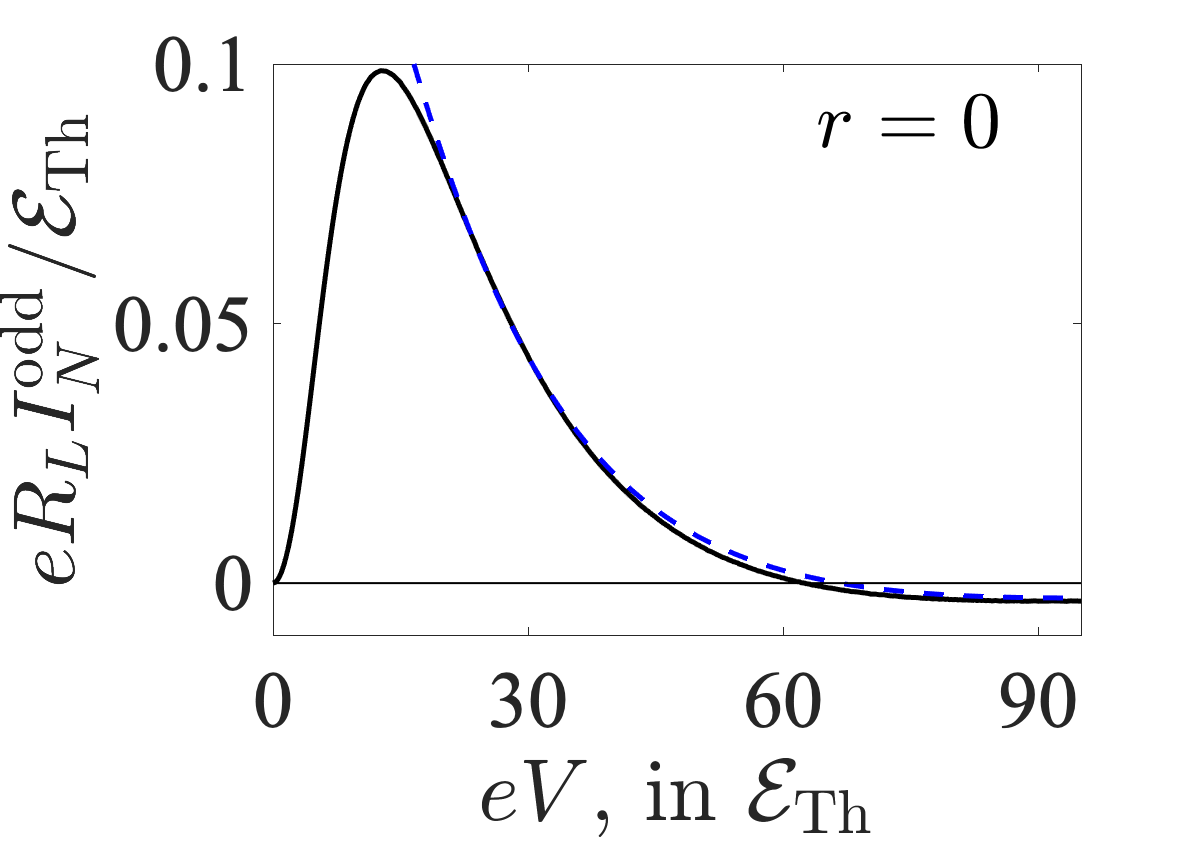}}
\def\stackalignment{l}
\topinset{\little}{\big}{-10pt}{105pt}\\
\end{minipage}
\vfill
\begin{minipage}[h]{1\linewidth}
\center{\includegraphics[width=1\linewidth]{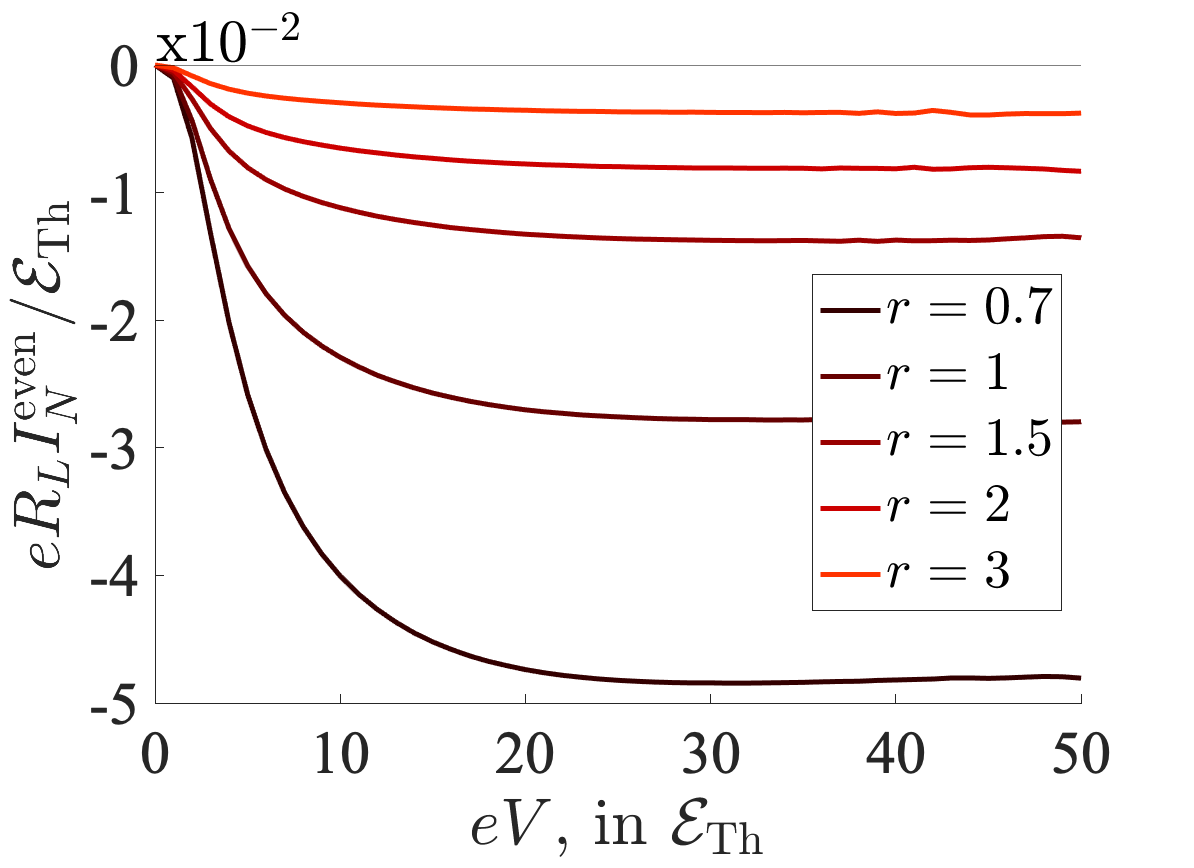}}
\end{minipage}
\caption{(Color online) Upper panel: The amplitude of the odd-in-$\phi$ oscillations of the current $I_N$ as a function of $V$ at $T\mapsto 0$ for different values of $r$. The parameter values are the same as in Fig.~\ref{fig:V1}, except $\Delta = 250{\cal E}_{\rm Th}$. Solid lines represent full numerical solution of the Usadel equation~(\ref{eqn:Usadel}), dashed lines correspond to the analytical result, cf. Eq.~(\ref{eqn::I_N_odd_full}). The latter equation provides a good approximation starting from $eV \gtrsim 20 {\cal E}_{\rm Th}$ down to the lowest values of $r$ employed in our calculation. Inset: The same quantity evaluated in the limit of fully transparent interfaces and $\Delta = 100{\cal E}_{\rm Th}$. Dashed line corresponds to Eq.~(\ref{eqn::I_N_odd_transparent}). Lower panel: Amplitude of the even-in-$\phi$ oscillations of $I_N$.
%%%%%%%%%%%%%%%%%%%%%%%%%%%%%%%%%%%%%%%%%%%%%%%%%% 
\label{fig::I_N}
%%%%%%%%%%%%%%%%%%%%%%%%%%%%%%%%%%%%%%%%%%%%%%%%%%
}
\end{figure}

Let us now turn to the electric current $I_N$ flowing between the two normal terminals. This current also demonstrates a $2\pi$-periodic dependence on phase $\phi$ and can be represented as a sum of even and odd 
in $\phi$ contributions:
\begin{eqnarray}
    I_N(\phi, V) = I^{\rm even}_N(\phi, V) + I^{\rm odd}_N(\phi, V).
\label{evenodd}
\end{eqnarray}
The first -- even in $\phi$ -- term again describes the Aharonov-Bohm-like contribution~\cite{C1} and is by no means surprising. At the same time, the presence of the odd periodic in $\phi$ contribution to current $I_N$ is curious. In contrast to the superconducting contour, here term $I^{\rm odd}_N(\phi, V)$ obviously cannot be attributed to the Josephson effect, and its physical nature requires further analysis.

For simplicity, let us assume that all cross sections are equal ${\cal A}_i = {\cal A}$. Then, we obtain
\begin{eqnarray}
    I^{\rm odd}_N = \frac{\sigma_N}{2e} \frac{\cal A}{l_{\rm N1} + l_{\rm N2}} \int d\epsilon j^{(0)}_L \int\limits_{-l_{\rm N2}}^{l_{\rm N1}} dx {\cal Y}(x,\epsilon),\label{eqn::I_odd}
\end{eqnarray}
where $j^{(0)}_L =(f_L^N(V_1) - f_L^N(V_2))/(l_{\rm N1} + l_{\rm N2})$ is the corresponding spectral current in the normal contour. Since the current  $I^{\rm odd}_N$ in Eq. (\ref{eqn::I_odd}) is controlled by the kinetic coefficient ${\cal Y}$ we conclude that this current should be attributed to the electron-hole asymmetry in our system generated due to the phase-sensitive mechanism of sequential Andreev reflections at different NS interfaces~\cite{kalenkov2017large}. This conclusion is further supported by observing that $I^{\rm odd}_N \propto \sin\phi/(r_Lr_R)$. 

At sufficiently large voltages $e|V|\gg {\cal E}_{\rm Th}$ the integrals in Eq. (\ref{eqn::I_odd}) can be handled explicitly, and we get
\begin{widetext}
\begin{eqnarray}
   \frac{eR_LI_N^{\rm odd}}{{\cal E}_{\rm Th}} &\approx& \frac{\sin\phi}{16 r_Rr_L} \frac{(R_{\rm S1} + R_{\rm S2} )^2}{(R_{\rm N1} + R_{\rm N2} )^2} \frac{L^2}{l_{\rm S1}^2 + l_{\rm S2}^2}  \sum_{i = 1,2} (-1)^{i+1} \frac{{\cal E}_{\rm Th}}{e|\Bar{V}_i|} \frac{\Delta^2 }{\Delta^2 - (e\Bar{V}_i)^2} e^{-\sqrt{e|\Bar{V}_i|/{\cal E}_{\rm Th}}} \times \notag{}\\
   &&\times \left[ \sin\left(\sqrt{\frac{e|\Bar{V}_i|}{{\cal E}_{\rm Th}}}\frac{l_{\rm S1} - l_{\rm S2}}{L}\right) + \frac{l_{\rm S1} - l_{\rm S2}}{L} \cos\left(\sqrt{\frac{e|\Bar{V}_i|}{{\cal E}_{\rm Th}}}\frac{l_{\rm S1} - l_{\rm S2}}{L}\right) \right], \label{eqn::I_N_odd_full}
\end{eqnarray}
\end{widetext}
where $R_{\textrm{S1}}=l_{\textrm{S1}}/(\mathcal{A}_{\textrm{S1}} \sigma_N)$ and $R_{\textrm{S2}}=l_{\textrm{S2}}/(\mathcal{A}_{\textrm{S2}} \sigma_N)$ are the normal state Drude resistances of the wires connected to the superconducting terminals.
We also note that the current in Eq.~(\ref{eqn::I_N_odd_full}) vanishes identically for $l_{\rm S1} = l_{\rm S2}$ (and $r_{\rm S1}=r_{\rm S2}$) and/or $l_{\rm N1} = l_{\rm N2}$, in full agreement with our symmetry considerations in Sec. IIC.

It is also interesting to study current $I_N(\phi, V)$ at different values of $r$. This can be done by numerically solving the Usadel equation~(\ref{eqn:Usadel}). In Fig.~\ref{fig::I_N} we  display the corresponding results for the amplitudes of both odd and even oscillations of the current $I_N$ as functions of $V$ at $T\mapsto 0$ for different values of $r$. We observe that the odd in $\phi$ harmonics persists at all values of $r$ becoming progressively more pronounced with decreasing $r$, cf. also the inset in Fig.~\ref{fig::I_N} where we present the result obtained in the limit $r=0$. This amplitude first increases with increasing $V$ reaching its maximum at around $eV \simeq 10 {\cal E}_{\rm Th}$ and then falls off being exponentially suppressed already at $eV \simeq 50 {\cal E}_{\rm Th}$ in accordance with Eq.~(\ref{eqn::I_N_odd_full}). Let us mention that in contrast to the current $I_J(V)$ which exhibits the transition to the $\pi$-junction state~\cite{dolgirev2018current, dolgirev2019interplay} at around $eV\gtrsim 15{\cal E}_{\rm Th}$ (cf. also Fig~\ref{fig::I_S}a), the amplitude of $I_N^{\rm odd}(V)$ demonstrates similar switching at much higher voltages $eV\gtrsim 60{\cal E}_{\rm Th}$. This behavior of $I_N^{\rm odd}(V)$ is related to the presence of an extra parameter $(l_{\rm S1} - l_{\rm S2})/L< 1$ in the argument of the $\sin$-term, cf. Eq.~(\ref{eqn::I_N_odd_full}).

The even in $\phi$ current amplitude $I_N^{\rm even} (V)$ saturates to a non-zero value at large voltages (see Fig.~\ref{fig::I_N}), just as one would expect for the Aharonov-Bohm-like contribution. Notably, the value of the plateau scales as $1/r^2$ for $r \gg 1$. This is in contrast to $I_{\rm AB}^{\infty}(r)$ which scales as $1/r^4$.

In order to complete our analysis of the current $I_N$, we note that with increasing temperature $T > eV$ the even in $\phi$ contribution to this current decays as a power law
similarly to $I_{\rm AB}(T)$, which could serve as a signature of the Aharonov-Bohm-like effect. In contrast, the odd in $\phi$ contribution $I_N^{\rm odd}(T)$ behaves qualitatively similarly to the Josephson term $I_{J}(T)$ decaying much faster than $I_N^{\rm even}(T)$ and becoming invisibly small already at temperatures of order  several ${\cal E}_{\rm Th}$.

\section{Conclusions}\label{sec:Conclusions}

In this work we performed a detailed analysis of a non-trivial interplay between proximity induced long-range quantum coherence and non-equilibrium effects in cross-like Andreev interferometers as well as of its impact on electron
transport properties of such devices. 

We emphasized a crucial role of various symmetries in our problem. The charge conjugation symmetry encoded in the Usadel equations allowed us to establish an important general relation (\ref{symcur}) which, in turn, helps to demonstrate that topology of  cross-like Andreev interferometers is essential for determining charge transport properties of these devices. 

We showed that in symmetric interferometers, the current $I_S$ in the superconducting contour is an odd function of the superconducting phase difference $\phi$. In other words, the even in $\phi$ Aharonov-Bohm-like contribution vanishes identically in such structures. These setups can only support the voltage-controlled Josephson current, and demonstrate switching between 0- and $\pi$-states depending on the applied voltage bias. In contrast, non-vanishing Aharonov-Bohm-like currents do survive in {\it asymmetric} structures. The physics of such devices is dominated by a trade-off between Josephson and Aharonov-Bohm-like quantum coherent contributions to the current $I_S$, leading to the $(I_0,\phi_0)$-junction state at sufficiently high bias voltages~\cite{dolgirev2018current,dolgirev2019interplay}. Hence, the current-phase relation $I_S(\phi)$ can be manipulated by external voltage bias, temperature and topology of the setup.

The current $I_N(\phi)$ flowing in the normal contour is also $2\pi$-periodic function of the superconducting phase difference $\phi$, i.e. it is directly affected by the proximity induced long-range quantum coherence. With the aid of our symmetry arguments we demonstrated that in symmetric Andreev interferometers, the current $I_N(\phi, V)$ is an even function of $\phi$ associated with the Aharonov-Bohm-like contribution.

A non-trivial effect discovered here is that in {\it asymmetric} cross-like interferometers the current $I_N$ develops an odd harmonics $I^{\rm odd}_N$, cf. Eq.~(\ref{eqn::I_odd}). The appearance of this contribution is particularly interesting because it can be attributed neither to the Aharonov-Bohm effect nor to the Josephson physics. In fact, our analysis demonstrates that the origin of the term $I^{\rm odd}_N$ is linked to violation of yet one more -- electron-hole -- symmetry that occurs under non-zero phase bias due to sequential Andreev reflections at different NS interfaces~\cite{kalenkov2017large}. In the tunneling limit the magnitude of this effect is controlled by $\sin \phi$ thereby resulting in a ``Josephson-like'' contribution  $I^{\rm odd}_N \propto \sin\phi$ to the current $I_N(\phi, V)$ between normal terminals. Similarly to $I_S(\phi)$, the current-phase relation $I_N(\phi)$ can also be manipulated by external voltage bias, temperature and topology of the interferometer.

Finally, we note that electron-hole symmetry violation is believed to also be responsible for large thermoelectric effects in Andreev interferometers~\cite{kalenkov2017large,dolgirev2019topology}. Our work, therefore, establishes an intimate relation between the current $I_N$ and thermoelectricity in hybrid superconducting nanostructures~\cite{KDZ}.

To conclude, we developed a detailed theory of quantum coherent charge transport in phase-and-voltage-biased asymmetric cross-like Andreev interferometers. The electron currents in both superconducting and normal contours demonstrate the presence of both even and odd $2\pi$-periodic in $\phi$ contributions. We identified key physical mechanisms responsible for different contributions to these currents, and described their non-trivial behavior depending on the applied voltage, temperature and the system topology. Our findings allow for full control of the current pattern in biased Andreev interferometers, thus rendering them particularly promising for future applications in modern electronics.

\vspace{0.5cm}

\centerline{\bf Acknowledgements}
The authors would like to thank A. Radkevich and A.G. Semenov for fruitful discussions. P.E.D. acknowledges the hospitality of Skoltech during the fall of 2018. P.E.D. and A.E.T. were supported by the Skoltech NGP program (Skoltech-MIT joint project). This work is partially supported by RFBR Grant No. 18-02-00586 for M.S.K. and A.D.Z.

\appendix
\section{Symmetric Andreev interferometers}
\label{appendix:Symmetry}

Let us focus our attention on the two special cases: (i) symmetric connectors to S-terminals with $l_{S1} = l_{S2}$ and $r_L=r_R$
and (ii) symmetric connectors to N-terminals with $l_{N1} = l_{N2}$.

In the case (i), we observe an extra symmetry related to the possibility of interchanging the terminals $S_1\leftrightarrow S_2$ with simultaneous inversion of the phase  $\phi \rightarrow -\phi$ implying that all the functions $V_1(\phi, V)$, $V_2(\phi, V)$, $I_{N1} (\phi, V_1, V_2)$, $I_{N2} (\phi, V_1, V_2)$ are even in $\phi$, cf. Eq. (\ref{ii}). Besides that, we have
\begin{gather}
I_{S1} (\phi, V) = - I_{S2} (-\phi, V),
\\
I_{S2} (\phi, V) = - I_{S1} (-\phi, V).
\end{gather}
Combining these equations with Eq.~\eqref{Scons} we arrive at the relation (\ref{i}).

In the case (ii), our system is symmetric with respect to interchanging the normal terminals $N_1\leftrightarrow N_2$. Then we have
\begin{gather}
I_{S1} (\phi, V_2, V_1) =  I_{S1} (\phi, V_1, V_2),
\label{NNS1}
\\
I_{S2} (\phi, V_2, V_1) = I_{S2} (\phi, V_1, V_2),
\label{NNS2}\\
I_{N1} (\phi, V_2, V_1) =  -I_{N2} (\phi, V_1, V_2),
\label{NNN1}
\\
I_{N2} (\phi, V_2, V_1) = -I_{N1} (\phi, V_1, V_2).
\label{NNN2}
\end{gather}

Let us define $\delta V=V_1+ V/2=V_2 - V/2$. By symmetry in the case (ii) the function $\delta V$ is even in $V$.
It follows from Eq.~\eqref{eqn:charge_conj} that for $\phi\rightarrow-\phi,\, V\rightarrow-V$ we get
\begin{eqnarray}
    &&I_{S1}(\phi,-V/2+\delta V(\phi)),V/2+\delta V(\phi)) =\notag{}\\
    &&-I_{S1}(-\phi,-(-V/2+\delta V(\phi)),-(V/2+\delta V(\phi)) =\notag{}\\
    &&-I_{S1}(-\phi,-V/2-\delta V(\phi),V/2-\delta V(\phi)).
\end{eqnarray}
Here we also employed Eq.~\eqref{NNS1}. Note that the currents $I_{S1},\, I_{S2}$ are even functions of $V$ and, hence, we obtain
\begin{equation}
\delta V(\phi, V) = - \delta V(-\phi, V),\label{deltaV}
\end{equation}
i.e. $\delta V$ turns out to be an odd function of the superconducting phase $\phi$. 

Applying the relations~\eqref{NNS1}, \eqref{deltaV} and \eqref{eqn:charge_conj} to the current $I_{S1}$ we get
\begin{multline}
I_{S1}(\phi, V_1(\phi), V_2(\phi)) = - I_{S1}(- \phi, - V_2(\phi), - V_1(\phi))
=\\=
- I_{S1}(- \phi,  V_1(-\phi),  V_2(-\phi)),
\end{multline}
implying that the current $I_{S}$ again turns out to be an odd function of the phase $\phi$, i.e. just like in the case (i) it
obeys the relation (\ref{i}).

Furthermore, making use of the relations \eqref{Ncons}, \eqref{NNN1}, \eqref{NNN2}, and \eqref{deltaV} we recover the following properties of the current $I_{N1}$:
\begin{multline}
I_{N1} (\phi, V_1(\phi), V_2(\phi)) = -I_{N2}(-\phi,-V_1(\phi),-V_2(\phi))\\=
I_{N1}(-\phi,-V_2(\phi),-V_1(\phi))=
I_{N1} (-\phi, V_1(-\phi), V_2(-\phi) ),
\end{multline}
implying that the current $I_{N}$ obeys the relation (\ref{ii}), thus being an even function of the phase $\phi$. 

Obviously, the relations (\ref{i}) and (\ref{ii}) are also obeyed in a special case of fully symmetric Andreev interferometers with $l_{S1} = l_{S2}$, $r_L = r_R$ and $l_{N1} = l_{N2}$. In this case we have $V_2 = - V_1 = V/2$, i.e. the potentials $V_1$ and $V_2$ do not depend on $\phi$.  

\section{Transparent SN interfaces}
While the main part of our paper is devoted to the tunneling limit described by KL boundary conditions  (\ref{eqn: Main_KL}),
it is useful to extend our analysis to the case of fully transparent SN interfaces corresponding to the limit $r=0$.
In particular, we performed a numerical analysis of the current $I_N$ (the current $I_S$ was investigated in Refs.~\cite{dolgirev2018current,dolgirev2019interplay}). The corresponding results are displayed in Fig.~\ref{fig::I_N} at different values of $r$, including $r=0$ in the inset of Fig.~\ref{fig::I_N}.

In the case of fully transparent SN interfaces and in the limit of high voltages $\mathcal{E}_{\mathrm{Th}} \ll e|V_{1,2}| < \Delta$ one can also derive an explicit expression for the odd harmonic of the current $I_N$. Employing the approach developed in Ref.~\cite{dolgirev2019interplay} one can easily find the anomalous Green's function in the normal wires connected to the normal terminals. We obtain
\begin{equation}
F^R = F_c^R e^{i \lambda x}, 
\quad
\tilde F^R = \tilde F_c^R e^{i \lambda x}, 
\end{equation}
where $x$ is the distance from the crossing point, and $F_c^R$ is the anomalous Green's function evaluated at this crossing point
\begin{multline}
F_c =
-\dfrac{8i \mathcal{A}_{S1} f_S e^{i \lambda l_{\mathrm{S1}}} e^{i\phi_L}}{
\mathcal{A}_{S1} + \mathcal{A}_{S2} + \mathcal{A}_{N1} + \mathcal{A}_{N2}
}
-\\-
\dfrac{8i \mathcal{A}_{S2} f_S e^{i \lambda l_{\mathrm{S2}}} e^{i\phi_R}}{
\mathcal{A}_{S1} + \mathcal{A}_{S2} + \mathcal{A}_{N1} + \mathcal{A}_{N2}
}.
\label{Fc}
\end{multline}
Here we defined 
\begin{equation}
f_S (\epsilon) = 
\tan\left[\dfrac{1}{4}\arcsin \dfrac{|\Delta|}{\sqrt{|\Delta|^2 - \epsilon^2}}\right].
\end{equation} 
The function $\tilde F_c$ can be recovered from Eq. \eqref{Fc} by replacing $F_c$ by $- \tilde F_c$ and $\chi_{1,2}$ by $-\chi_{1,2}$. Then it is straightforward to derive the function $\mathcal{Y}$ and evaluate the integral in Eq.~\eqref{eqn::I_odd}. In the case of equal cross sections we obtain
\begin{widetext}
\begin{eqnarray}
    \frac{eR_LI_N^{\rm odd}}{{\cal E}_{\rm Th}} &\approx& \frac{4 \sin\phi}{l_{\rm N1} + l_{\rm N2}} \frac{L^2}{l_{\rm S1}^2+l_{\rm S2}^2} \sum_{i = 1,2} (-1)^{i+1} f_S^2(e|V_i|) \exp\left( -\sqrt{\frac{e|V_i|}{{\cal E}_{\rm Th}}} \right)\notag{}\\
    &\times&\left[ L \sin{\left( \frac{l_{\rm S1} - l_{\rm S2}}{L}\sqrt{\frac{e|V_i|}{{\cal E}_{\rm Th}}} \right)} + (l_{\rm S1} - l_{\rm S2}) \cos{\left( \frac{l_{\rm S1} - l_{\rm S2}}{L}\sqrt{\frac{e|V_i|}{{\cal E}_{\rm Th}}} \right)} \right]. \label{eqn::I_N_odd_transparent}
\end{eqnarray}
\end{widetext}
As shown in the inset of Fig.~\ref{fig::I_N}, this analytic expression perfectly matches with our numerical result at  $eV \gtrsim 20{\cal E}_{\rm Th}$. In addition, Eq.~(\ref{eqn::I_N_odd_transparent}) can be used to estimate the maximum value of the current $I^{\rm odd}_N$.

\bibliography{Andreev_interferometers}

\begin{thebibliography}{37}
\expandafter\ifx\csname natexlab\endcsname\relax\def\natexlab#1{#1}\fi
\expandafter\ifx\csname bibnamefont\endcsname\relax
  \def\bibnamefont#1{#1}\fi
\expandafter\ifx\csname bibfnamefont\endcsname\relax
  \def\bibfnamefont#1{#1}\fi
\expandafter\ifx\csname citenamefont\endcsname\relax
  \def\citenamefont#1{#1}\fi
\expandafter\ifx\csname url\endcsname\relax
  \def\url#1{\texttt{#1}}\fi
\expandafter\ifx\csname urlprefix\endcsname\relax\def\urlprefix{URL }\fi
\providecommand{\bibinfo}[2]{#2}
\providecommand{\eprint}[2][]{\url{#2}}

\bibitem[{\citenamefont{Belzig et~al.}(1999)\citenamefont{Belzig, Wilhelm,
  Bruder, Sch{\"o}n, and Zaikin}}]{belzig1999quasiclassical}
\bibinfo{author}{\bibfnamefont{W.}~\bibnamefont{Belzig}},
  \bibinfo{author}{\bibfnamefont{F.~K.} \bibnamefont{Wilhelm}},
  \bibinfo{author}{\bibfnamefont{C.}~\bibnamefont{Bruder}},
  \bibinfo{author}{\bibfnamefont{G.}~\bibnamefont{Sch{\"o}n}},
  \bibnamefont{and} \bibinfo{author}{\bibfnamefont{A.~D.}
  \bibnamefont{Zaikin}}, \bibinfo{journal}{Superlatt. Microstruct.}
  \textbf{\bibinfo{volume}{25}}, \bibinfo{pages}{1251} (\bibinfo{year}{1999}).

\bibitem[{\citenamefont{Fornieri and Giazotto}(2017)}]{fornieri2017towards}
\bibinfo{author}{\bibfnamefont{A.}~\bibnamefont{Fornieri}} \bibnamefont{and}
  \bibinfo{author}{\bibfnamefont{F.}~\bibnamefont{Giazotto}},
  \bibinfo{journal}{Nat. Nanotech.} \textbf{\bibinfo{volume}{12}},
  \bibinfo{pages}{944} (\bibinfo{year}{2017}).

\bibitem[{\citenamefont{Giazotto et~al.}(2006)\citenamefont{Giazotto,
  Heikkil{\"a}, Luukanen, Savin, and Pekola}}]{giazotto2006opportunities}
\bibinfo{author}{\bibfnamefont{F.}~\bibnamefont{Giazotto}},
  \bibinfo{author}{\bibfnamefont{T.~T.} \bibnamefont{Heikkil{\"a}}},
  \bibinfo{author}{\bibfnamefont{A.}~\bibnamefont{Luukanen}},
  \bibinfo{author}{\bibfnamefont{A.~M.} \bibnamefont{Savin}}, \bibnamefont{and}
  \bibinfo{author}{\bibfnamefont{J.~P.} \bibnamefont{Pekola}},
  \bibinfo{journal}{Rev. Mod. Phys.} \textbf{\bibinfo{volume}{78}},
  \bibinfo{pages}{217} (\bibinfo{year}{2006}).

\bibitem[{\citenamefont{Eom et~al.}(1998)\citenamefont{Eom, Chien, and
  Chandrasekhar}}]{eom1998phase}
\bibinfo{author}{\bibfnamefont{J.}~\bibnamefont{Eom}},
  \bibinfo{author}{\bibfnamefont{C.-J.} \bibnamefont{Chien}}, \bibnamefont{and}
  \bibinfo{author}{\bibfnamefont{V.}~\bibnamefont{Chandrasekhar}},
  \bibinfo{journal}{Phys. Rev. Lett.} \textbf{\bibinfo{volume}{81}},
  \bibinfo{pages}{437} (\bibinfo{year}{1998}).

\bibitem[{\citenamefont{Dikin et~al.}(2001)\citenamefont{Dikin, Jung, and
  Chandrasekhar}}]{dikin2001low}
\bibinfo{author}{\bibfnamefont{D.~A.} \bibnamefont{Dikin}},
  \bibinfo{author}{\bibfnamefont{S.}~\bibnamefont{Jung}}, \bibnamefont{and}
  \bibinfo{author}{\bibfnamefont{V.}~\bibnamefont{Chandrasekhar}},
  \bibinfo{journal}{Phys. Rev. B} \textbf{\bibinfo{volume}{65}},
  \bibinfo{pages}{012511} (\bibinfo{year}{2001}).

\bibitem[{\citenamefont{Parsons et~al.}(2003)\citenamefont{Parsons, Sosnin, and
  Petrashov}}]{parsons2003reversal}
\bibinfo{author}{\bibfnamefont{A.}~\bibnamefont{Parsons}},
  \bibinfo{author}{\bibfnamefont{I.~A.} \bibnamefont{Sosnin}},
  \bibnamefont{and} \bibinfo{author}{\bibfnamefont{V.~T.}
  \bibnamefont{Petrashov}}, \bibinfo{journal}{Phys. Rev. B}
  \textbf{\bibinfo{volume}{67}}, \bibinfo{pages}{140502}
  (\bibinfo{year}{2003}).

\bibitem[{\citenamefont{Cadden-Zimansky
  et~al.}(2007)\citenamefont{Cadden-Zimansky, Jiang, and
  Chandrasekhar}}]{cadden2007charge}
\bibinfo{author}{\bibfnamefont{P.}~\bibnamefont{Cadden-Zimansky}},
  \bibinfo{author}{\bibfnamefont{Z.}~\bibnamefont{Jiang}}, \bibnamefont{and}
  \bibinfo{author}{\bibfnamefont{V.}~\bibnamefont{Chandrasekhar}},
  \bibinfo{journal}{New J. Phys.} \textbf{\bibinfo{volume}{9}},
  \bibinfo{pages}{116} (\bibinfo{year}{2007}).

\bibitem[{\citenamefont{Shelly et~al.}(2016)\citenamefont{Shelly, Matrozova,
  and Petrashov}}]{shelly2016resolving}
\bibinfo{author}{\bibfnamefont{C.~D.} \bibnamefont{Shelly}},
  \bibinfo{author}{\bibfnamefont{E.~A.} \bibnamefont{Matrozova}},
  \bibnamefont{and} \bibinfo{author}{\bibfnamefont{V.~T.}
  \bibnamefont{Petrashov}}, \bibinfo{journal}{Sci. Adv.}
  \textbf{\bibinfo{volume}{2}}, \bibinfo{pages}{e1501250}
  (\bibinfo{year}{2016}).

\bibitem[{\citenamefont{Stoof and Nazarov}(1996)}]{stoof1996flux}
\bibinfo{author}{\bibfnamefont{T.~H.} \bibnamefont{Stoof}} \bibnamefont{and}
  \bibinfo{author}{\bibfnamefont{Y.~V.} \bibnamefont{Nazarov}},
  \bibinfo{journal}{Phys. Rev. B} \textbf{\bibinfo{volume}{54}},
  \bibinfo{pages}{R772} (\bibinfo{year}{1996}).

\bibitem[{\citenamefont{Golubov et~al.}(1997)\citenamefont{Golubov, Wilhelm,
  and Zaikin}}]{golubov1997coherent}
\bibinfo{author}{\bibfnamefont{A.~A.} \bibnamefont{Golubov}},
  \bibinfo{author}{\bibfnamefont{F.~K.} \bibnamefont{Wilhelm}},
  \bibnamefont{and} \bibinfo{author}{\bibfnamefont{A.~D.}
  \bibnamefont{Zaikin}}, \bibinfo{journal}{Phys. Rev. B}
  \textbf{\bibinfo{volume}{55}}, \bibinfo{pages}{1123} (\bibinfo{year}{1997}).

\bibitem[{\citenamefont{Nakano and Takayanagi}(1991)}]{nakano1991quasiparticle}
\bibinfo{author}{\bibfnamefont{H.}~\bibnamefont{Nakano}} \bibnamefont{and}
  \bibinfo{author}{\bibfnamefont{H.}~\bibnamefont{Takayanagi}},
  \bibinfo{journal}{Solid State Commun.} \textbf{\bibinfo{volume}{80}},
  \bibinfo{pages}{997} (\bibinfo{year}{1991}).

\bibitem[{\citenamefont{Petrashov et~al.}(1995)\citenamefont{Petrashov,
  Antonov, Delsing, and Claeson}}]{petrashov1995phase}
\bibinfo{author}{\bibfnamefont{V.~T.} \bibnamefont{Petrashov}},
  \bibinfo{author}{\bibfnamefont{V.~N.} \bibnamefont{Antonov}},
  \bibinfo{author}{\bibfnamefont{P.}~\bibnamefont{Delsing}}, \bibnamefont{and}
  \bibinfo{author}{\bibfnamefont{T.}~\bibnamefont{Claeson}},
  \bibinfo{journal}{Phys. Rev. Lett.} \textbf{\bibinfo{volume}{74}},
  \bibinfo{pages}{5268} (\bibinfo{year}{1995}).

\bibitem[{\citenamefont{Courtois et~al.}(1996)\citenamefont{Courtois, Gandit,
  Mailly, and Pannetier}}]{courtois1996long}
\bibinfo{author}{\bibfnamefont{H.}~\bibnamefont{Courtois}},
  \bibinfo{author}{\bibfnamefont{P.}~\bibnamefont{Gandit}},
  \bibinfo{author}{\bibfnamefont{D.}~\bibnamefont{Mailly}}, \bibnamefont{and}
  \bibinfo{author}{\bibfnamefont{B.}~\bibnamefont{Pannetier}},
  \bibinfo{journal}{Phys. Rev. Lett.} \textbf{\bibinfo{volume}{76}},
  \bibinfo{pages}{130} (\bibinfo{year}{1996}).

\bibitem[{\citenamefont{Byers and Flatt{\'e}}(1995)}]{byers1995probing}
\bibinfo{author}{\bibfnamefont{J.~M.} \bibnamefont{Byers}} \bibnamefont{and}
  \bibinfo{author}{\bibfnamefont{M.~E.} \bibnamefont{Flatt{\'e}}},
  \bibinfo{journal}{Phys. Rev. Lett.} \textbf{\bibinfo{volume}{74}},
  \bibinfo{pages}{306} (\bibinfo{year}{1995}).

\bibitem[{\citenamefont{Deutscher and Feinberg}(2000)}]{deutscher2000coupling}
\bibinfo{author}{\bibfnamefont{G.}~\bibnamefont{Deutscher}} \bibnamefont{and}
  \bibinfo{author}{\bibfnamefont{D.}~\bibnamefont{Feinberg}},
  \bibinfo{journal}{Appl. Phys. Lett.} \textbf{\bibinfo{volume}{76}},
  \bibinfo{pages}{487} (\bibinfo{year}{2000}).

\bibitem[{\citenamefont{Beckmann et~al.}(2004)\citenamefont{Beckmann, Weber,
  and L{\"o}hneysen}}]{beckmann2004evidence}
\bibinfo{author}{\bibfnamefont{D.}~\bibnamefont{Beckmann}},
  \bibinfo{author}{\bibfnamefont{H.~B.} \bibnamefont{Weber}}, \bibnamefont{and}
  \bibinfo{author}{\bibfnamefont{H.~v.} \bibnamefont{L{\"o}hneysen}},
  \bibinfo{journal}{Phys. Rev. Lett.} \textbf{\bibinfo{volume}{93}},
  \bibinfo{pages}{197003} (\bibinfo{year}{2004}).

\bibitem[{\citenamefont{Russo et~al.}(2005)\citenamefont{Russo, Kroug,
  Klapwijk, and Morpurgo}}]{russo2005experimental}
\bibinfo{author}{\bibfnamefont{S.}~\bibnamefont{Russo}},
  \bibinfo{author}{\bibfnamefont{M.}~\bibnamefont{Kroug}},
  \bibinfo{author}{\bibfnamefont{T.~M.} \bibnamefont{Klapwijk}},
  \bibnamefont{and} \bibinfo{author}{\bibfnamefont{A.~F.}
  \bibnamefont{Morpurgo}}, \bibinfo{journal}{Phys. Rev. Lett.}
  \textbf{\bibinfo{volume}{95}}, \bibinfo{pages}{027002}
  (\bibinfo{year}{2005}).

\bibitem[{\citenamefont{Cadden-Zimansky and
  Chandrasekhar}(2006)}]{cadden2006nonlocal}
\bibinfo{author}{\bibfnamefont{P.}~\bibnamefont{Cadden-Zimansky}}
  \bibnamefont{and}
  \bibinfo{author}{\bibfnamefont{V.}~\bibnamefont{Chandrasekhar}},
  \bibinfo{journal}{Phys. Rev. Lett.} \textbf{\bibinfo{volume}{97}},
  \bibinfo{pages}{237003} (\bibinfo{year}{2006}).

\bibitem[{\citenamefont{Golubev and Zaikin}(2007)}]{golubev2007non}
\bibinfo{author}{\bibfnamefont{D.~S.} \bibnamefont{Golubev}} \bibnamefont{and}
  \bibinfo{author}{\bibfnamefont{A.~D.} \bibnamefont{Zaikin}},
  \bibinfo{journal}{Phys. Rev. B} \textbf{\bibinfo{volume}{76}},
  \bibinfo{pages}{184510} (\bibinfo{year}{2007}).

\bibitem[{\citenamefont{Kalenkov and Zaikin}(2007)}]{kalenkov2007crossed}
\bibinfo{author}{\bibfnamefont{M.~S.} \bibnamefont{Kalenkov}} \bibnamefont{and}
  \bibinfo{author}{\bibfnamefont{A.~D.} \bibnamefont{Zaikin}},
  \bibinfo{journal}{Phys. Rev. B} \textbf{\bibinfo{volume}{76}},
  \bibinfo{pages}{224506} (\bibinfo{year}{2007}).

\bibitem[{\citenamefont{Golubev et~al.}(2009)\citenamefont{Golubev, Kalenkov,
  and Zaikin}}]{golubev2009crossed}
\bibinfo{author}{\bibfnamefont{D.~S.} \bibnamefont{Golubev}},
  \bibinfo{author}{\bibfnamefont{M.~S.} \bibnamefont{Kalenkov}},
  \bibnamefont{and} \bibinfo{author}{\bibfnamefont{A.~D.}
  \bibnamefont{Zaikin}}, \bibinfo{journal}{Phys. Rev. Lett.}
  \textbf{\bibinfo{volume}{103}}, \bibinfo{pages}{067006}
  (\bibinfo{year}{2009}).

\bibitem[{ZZh()}]{ZZh}
\bibinfo{note}{A.D. Zaikin and G.F. Zharkov, Fiz. Nizk. Temp. {\bf 7}, 375
  (1981) [Sov. J. Low Temp. Phys. {\bf 7}, 181 (1981)].}

\bibitem[{\citenamefont{Dubos et~al.}(2001)\citenamefont{Dubos, Courtois,
  Pannetier, Wilhelm, Zaikin, and Sch{\"o}n}}]{dubos2001josephson}
\bibinfo{author}{\bibfnamefont{P.}~\bibnamefont{Dubos}},
  \bibinfo{author}{\bibfnamefont{H.}~\bibnamefont{Courtois}},
  \bibinfo{author}{\bibfnamefont{B.}~\bibnamefont{Pannetier}},
  \bibinfo{author}{\bibfnamefont{F.}~\bibnamefont{Wilhelm}},
  \bibinfo{author}{\bibfnamefont{A.}~\bibnamefont{Zaikin}}, \bibnamefont{and}
  \bibinfo{author}{\bibfnamefont{G.}~\bibnamefont{Sch{\"o}n}},
  \bibinfo{journal}{Phys. Rev. B} \textbf{\bibinfo{volume}{63}},
  \bibinfo{pages}{064502} (\bibinfo{year}{2001}).

\bibitem[{\citenamefont{Golubov et~al.}(2004)\citenamefont{Golubov, Kupriyanov,
  and Il’Ichev}}]{golubov2004current}
\bibinfo{author}{\bibfnamefont{A.~A.} \bibnamefont{Golubov}},
  \bibinfo{author}{\bibfnamefont{M.~Y.} \bibnamefont{Kupriyanov}},
  \bibnamefont{and}
  \bibinfo{author}{\bibfnamefont{E.}~\bibnamefont{Il’Ichev}},
  \bibinfo{journal}{Rev. Mod. Phys.} \textbf{\bibinfo{volume}{76}},
  \bibinfo{pages}{411} (\bibinfo{year}{2004}).

\bibitem[{\citenamefont{Volkov}(1995)}]{volkov1995af}
\bibinfo{author}{\bibfnamefont{A.~F.} \bibnamefont{Volkov}},
  \bibinfo{journal}{Phys. Rev. Lett.} \textbf{\bibinfo{volume}{74}},
  \bibinfo{pages}{4730} (\bibinfo{year}{1995}).

\bibitem[{\citenamefont{Wilhelm et~al.}(1998)\citenamefont{Wilhelm, Sch{\"o}n,
  and Zaikin}}]{wilhelm1998mesoscopic}
\bibinfo{author}{\bibfnamefont{F.~K.} \bibnamefont{Wilhelm}},
  \bibinfo{author}{\bibfnamefont{G.}~\bibnamefont{Sch{\"o}n}},
  \bibnamefont{and} \bibinfo{author}{\bibfnamefont{A.~D.}
  \bibnamefont{Zaikin}}, \bibinfo{journal}{Phys. Rev. Lett.}
  \textbf{\bibinfo{volume}{81}}, \bibinfo{pages}{1682} (\bibinfo{year}{1998}).

\bibitem[{\citenamefont{Yip}(1998)}]{yip1998energy}
\bibinfo{author}{\bibfnamefont{S.-K.} \bibnamefont{Yip}},
  \bibinfo{journal}{Phys. Rev. B} \textbf{\bibinfo{volume}{58}},
  \bibinfo{pages}{5803} (\bibinfo{year}{1998}).

\bibitem[{\citenamefont{Baselmans et~al.}(1999)\citenamefont{Baselmans,
  Morpurgo, Van~Wees, and Klapwijk}}]{baselmans1999reversing}
\bibinfo{author}{\bibfnamefont{J.~J.~A.} \bibnamefont{Baselmans}},
  \bibinfo{author}{\bibfnamefont{A.~F.} \bibnamefont{Morpurgo}},
  \bibinfo{author}{\bibfnamefont{B.~J.} \bibnamefont{Van~Wees}},
  \bibnamefont{and} \bibinfo{author}{\bibfnamefont{T.~M.}
  \bibnamefont{Klapwijk}}, \bibinfo{journal}{Nature}
  \textbf{\bibinfo{volume}{397}}, \bibinfo{pages}{43} (\bibinfo{year}{1999}).

\bibitem[{\citenamefont{Dolgirev et~al.}(2018)\citenamefont{Dolgirev, Kalenkov,
  and Zaikin}}]{dolgirev2018current}
\bibinfo{author}{\bibfnamefont{P.~E.} \bibnamefont{Dolgirev}},
  \bibinfo{author}{\bibfnamefont{M.~S.} \bibnamefont{Kalenkov}},
  \bibnamefont{and} \bibinfo{author}{\bibfnamefont{A.~D.}
  \bibnamefont{Zaikin}}, \bibinfo{journal}{Phys. Rev. B}
  \textbf{\bibinfo{volume}{97}}, \bibinfo{pages}{054521}
  (\bibinfo{year}{2018}).

\bibitem[{\citenamefont{Dolgirev
  et~al.}(2019{\natexlab{a}})\citenamefont{Dolgirev, Kalenkov, and
  Zaikin}}]{dolgirev2019interplay}
\bibinfo{author}{\bibfnamefont{P.~E.} \bibnamefont{Dolgirev}},
  \bibinfo{author}{\bibfnamefont{M.~S.} \bibnamefont{Kalenkov}},
  \bibnamefont{and} \bibinfo{author}{\bibfnamefont{A.~D.}
  \bibnamefont{Zaikin}}, \bibinfo{journal}{Sci. Rep.}
  \textbf{\bibinfo{volume}{9}}, \bibinfo{pages}{1301}
  (\bibinfo{year}{2019}{\natexlab{a}}).

\bibitem[{\citenamefont{Kalenkov and Zaikin}(2017)}]{kalenkov2017large}
\bibinfo{author}{\bibfnamefont{M.~S.} \bibnamefont{Kalenkov}} \bibnamefont{and}
  \bibinfo{author}{\bibfnamefont{A.~D.} \bibnamefont{Zaikin}},
  \bibinfo{journal}{Phys. Rev. B} \textbf{\bibinfo{volume}{95}},
  \bibinfo{pages}{024518} (\bibinfo{year}{2017}).

\bibitem[{\citenamefont{Schopohl and Maki}(1995)}]{schopohl1995quasiparticle}
\bibinfo{author}{\bibfnamefont{N.}~\bibnamefont{Schopohl}} \bibnamefont{and}
  \bibinfo{author}{\bibfnamefont{K.}~\bibnamefont{Maki}},
  \bibinfo{journal}{Phys. Rev. B} \textbf{\bibinfo{volume}{52}},
  \bibinfo{pages}{490} (\bibinfo{year}{1995}).

\bibitem[{\citenamefont{Schopohl}(1998)}]{schopohl1998transformation}
\bibinfo{author}{\bibfnamefont{N.}~\bibnamefont{Schopohl}},
  \bibinfo{journal}{arXiv eprint:cond-mat/9804064}  (\bibinfo{year}{1998}).

\bibitem[{\citenamefont{Kuprianov and Lukichev}(1988)}]{kuprianov1988influence}
\bibinfo{author}{\bibfnamefont{M.~Y.} \bibnamefont{Kuprianov}}
  \bibnamefont{and} \bibinfo{author}{\bibfnamefont{V.~F.}
  \bibnamefont{Lukichev}}, \bibinfo{journal}{Zh. Eksp. Teor. Fiz}
  \textbf{\bibinfo{volume}{94}}, \bibinfo{pages}{149} (\bibinfo{year}{1988}).

\bibitem[{C1()}]{C1}
\bibinfo{note}{Taking advantage of the normalization condition $\check{G}^2 =
  \check{1}$ it becomes possible to derive the expression for $I_N$ to the
  first order in $1/r^2$ without expanding the spectral functions
  $\gamma,\tilde{\gamma}$ to the next order.}

\bibitem[{\citenamefont{Dolgirev
  et~al.}(2019{\natexlab{b}})\citenamefont{Dolgirev, Kalenkov, and
  Zaikin}}]{dolgirev2019topology}
\bibinfo{author}{\bibfnamefont{P.~E.} \bibnamefont{Dolgirev}},
  \bibinfo{author}{\bibfnamefont{M.~S.} \bibnamefont{Kalenkov}},
  \bibnamefont{and} \bibinfo{author}{\bibfnamefont{A.~D.}
  \bibnamefont{Zaikin}}, \bibinfo{journal}{Physica Status Solidi (RRL)}
  \textbf{\bibinfo{volume}{13}}, \bibinfo{pages}{1800252}
  (\bibinfo{year}{2019}{\natexlab{b}}).

\bibitem[{KDZ()}]{KDZ}
\bibinfo{note}{M. S. Kalenkov, P. E. Dolgirev, and A. D. Zaikin, in
  preparation.}

\end{thebibliography}

\end{document}